\documentclass[twocolumn,pre,aps,showpacs,a4paper,floatfix,amssymb]{revtex4-1}

\usepackage{graphicx}
\usepackage{psfrag}
\usepackage{subfigure}
\usepackage{epsfig}
\usepackage{calc}
\usepackage{bm}
\usepackage{longtable}
\usepackage{ulem}
\usepackage{color}
\graphicspath{{figures/}}
\usepackage{hyperref}
\hypersetup{
  colorlinks,
  citecolor=[rgb]{0,0,1},
  linkcolor=[rgb]{0,0,1},
  urlcolor=[rgb]{0,0,1}}
  
\newcommand{\ud}{d}
\newcommand{\rod}{\mathrm{rod}}
\newcommand{\gas}{\mathrm{gas}}
\newcommand{\Jatt}{J_\mathrm{att}}
\newcommand{\Jdet}{J_\mathrm{det}}
\newcommand{\kBT}{k_\mathrm{B}T}
\newcommand{\Pe}{\mathrm{Pe}}

\begin{document}

\title{Collective behavior of penetrable self-propelled rods in two dimensions}
\author{Masoud Abkenar, Kristian Marx, Thorsten Auth, and Gerhard Gompper}
\affiliation{Theoretical Soft Matter and Biophysics, Institute of Complex Systems and
 Institute for Advanced Simulation, Forschungszentrum J\"ulich, D-52425 J\"ulich, Germany}


\begin{abstract}

Collective behavior of self-propelled particles is observed on a microscale for swimmers such as sperm and bacteria as well as for protein filaments in motility assays.
The properties of such systems depend both on their dimensionality and the interactions between their particles. 
We introduce a model for self-propelled rods in two dimensions that interact via a separation-shifted Lennard-Jones potential. Due to the finite potential barrier, the rods are able to cross. 
This model allows us to efficiently simulate systems of self-propelled rods that effectively move in two dimensions but can occasionally escape to the third dimension in order to pass each other.
Our quasi-two-dimensional self-propelled particles describe a class of active systems that encompasses microswimmers close to a wall and filaments propelled on a substrate. 
Using Monte Carlo simulations, we first determine the isotropic-nematic transition for passive rods. 
Using Brownian dynamics simulations, we characterize cluster formation of self-propelled rods as a function of propulsion strength, noise, and energy barrier. 
Contrary to rods with an infinite potential barrier, an increase of the propulsion strength does not only favor alignment but also effectively decreases the potential barrier that prevents crossing of rods. 
We thus find a clustering window with a maximum cluster size at medium propulsion strengths. 
\end{abstract}

\maketitle

\section{Introduction}
Collective behavior of active bodies is frequently found in macroscopic systems such as bird flocks and fish schools \cite{Vicsek1208}, but also is found in microscopic systems such as sperm cells \cite{Riedel2005,Yang2008}, bacteria \cite{BenJacob2000,Sokolov2007,Peruani2012,Gachelin2013}, and 
manmade microswimmers that propel themselves forward using a chemical or physical mechanism \cite{Paxton2004,Volpe2011,Mei2011,Ebbens2012,Yang2011}.
Despite the different natures of these systems, they all exhibit interactions that favor alignment of neighboring bodies, thus leading to similar forms of collective behavior.
Of particular interest for us are experiments with elongated self-propelled
particles on the microscopic scale in two dimensions, such as motility
assays where actin filaments are propelled on a carpet of myosin motor
proteins \cite{Harada1987,Schaller2010}, microtubules propelled by surface-bound dyneins
\cite{Sumino2012}, and microswimmers that are attracted to surfaces \cite{lauga2006,Berke2008,Elgeti2009,Elgeti2010,Elgeti1302}.

In the pioneering work of Vicsek \textit{et al.} \cite{Vicsek1995}, 
nonequilibrium phase transitions were observed for systems with self-propelled point particles that interact via an imposed alignment rule and thermal noise. 
This work led to numerous analytical \cite{Toner1995,Simha2002,Ramaswamy2003,Toner2005,Peruani2008,Bertin2009,Golestanian2009} 
as well as computational \cite{redner1301, Szabo2006,Gregoire2004,Huepe2004,Dorsogna2006,Aldana2007,
Chate2008,Ginelli2010} 
studies for systems of self-propelled particles.
Because each particle consumes energy to generate motion, the systems are far from equilibrium and interesting new dynamic properties emerge.
For rods with strong short-range repulsive interactions (volume exclusion) it has been shown that self-propelled motion leads to alignment of rods \cite{Baskaran2008b,Mccandlish2012,Baskaran2008,Yang2010,Wensink2012,Kraikivski2006,costanzo1201}. 
Moreover, self-propulsion enhances aggregation and cluster formation \cite{Peruani2006,Wensink2012,Ginelli2010,Kraikivski2006,Peruani2010,Yang2010}. 
Near the transition from a disordered to an ordered state, the cluster size distribution obeys a power-law decay \cite{Huepe2004,Huepe2008,Yang2010,Peruani2012}. 
In simulations at higher densities, longitudinally moving bands \cite{Ginelli2010} and lanes \cite{Mccandlish2012,Wensink2012} have been observed.

Motility assays with actin filaments or microtubules are essentially two-dimensional systems, but with a finite probability for the filaments to cross each other \cite{Ruhnow2011,Sumino2012}. 
Because the filaments are not tightly bound to the surface, one of them might
be slightly and temporarily pushed away from the surface when two filaments collide. 
In Ref.~\cite{Sumino2012}, microtubules have been found to cross each other with a probability of $40 \, \%$ if they approach perpendicularly. 
Two-dimensional models with impenetrable swimmers thus do not adequately describe these systems, while full three-dimensional calculations are computationally expensive. 
In Ref.~\cite{Schaller2010}, a cellular automaton model with an imposed
alignment rule that allows two filaments to occupy the same site has been
used to simulate actin motility assays. 

In this paper, we propose a model for self-propelled rods (SPRs) in two dimensions
that interact with a physical interaction potential.
We discretize each rod by a number of beads to calculate rod-rod interactions.
In contrast to previous models with strict excluded-volume interactions  \cite{Peruani2006,Yang2010,Wensink2012,Mccandlish2012,Kraikivski2006,Peruani2006}, our interaction potential allows rods to cross.
Our simulations thus combine the computational efficiency of
two-dimensional simulations with a possibility to mimic an escape to the third
dimension when two rods collide.
Simulation snapshots of the system 
which display disordered states, motile clusters, lanes, etc. 
are shown in Fig.~\ref{fig:snapshots}, and movies can be found in the Supplemental Material \cite{Abkenar13suppl}.

The paper is organized as follows. 
We introduce model, simulation methods, and numerical parameters in Sec.~\ref{sec:model}.
We calculate a phase diagram for passive (nonswimming) rods in Sec.~\ref{sec:passiverods} using Monte Carlo simulations, followed by a short discussion on the probability of crossing events in Sec.~\ref{sec:crossing}.
We focus on cluster formation in Sec.~\ref{sec:clusterformation}, introducing gas density and cluster break-up in Sec.~\ref{sec:voiddensity}, cluster size analysis in Sec.~\ref{sec:csd}, and autocorrelation functions for rod orientations in Sec.~\ref{sec:oacf}. 
We summarize our main results in Sec.~\ref{sec:summary}.

\begin{figure*}[t]
\centering
  \subfigure[~$\Pe=0,\, \rho\,L_\rod^2=10.2$]{\label{snapshot_a}\includegraphics[totalheight=0.5\columnwidth]{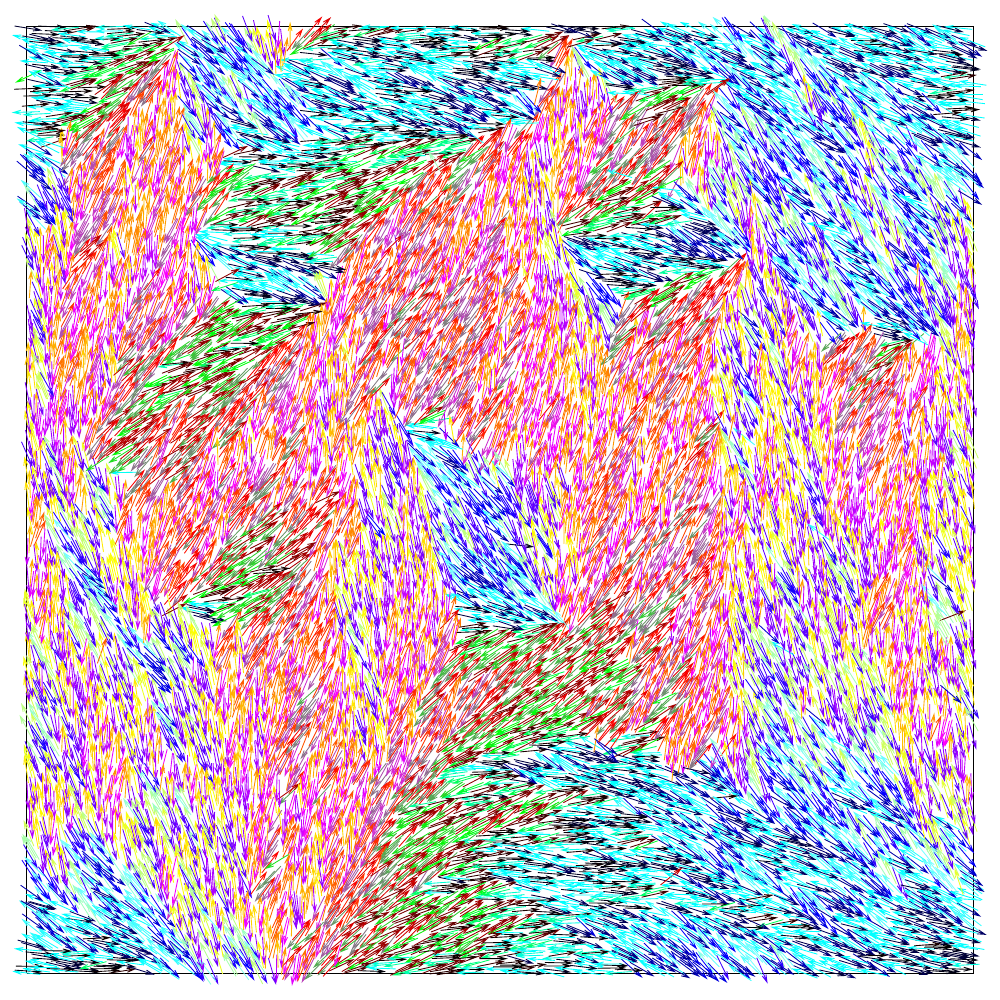}}
  \subfigure[~$\Pe=0,\, \rho\,L_\rod^2=5.1$]{\label{snapshot_b}\includegraphics[totalheight=0.5\columnwidth]{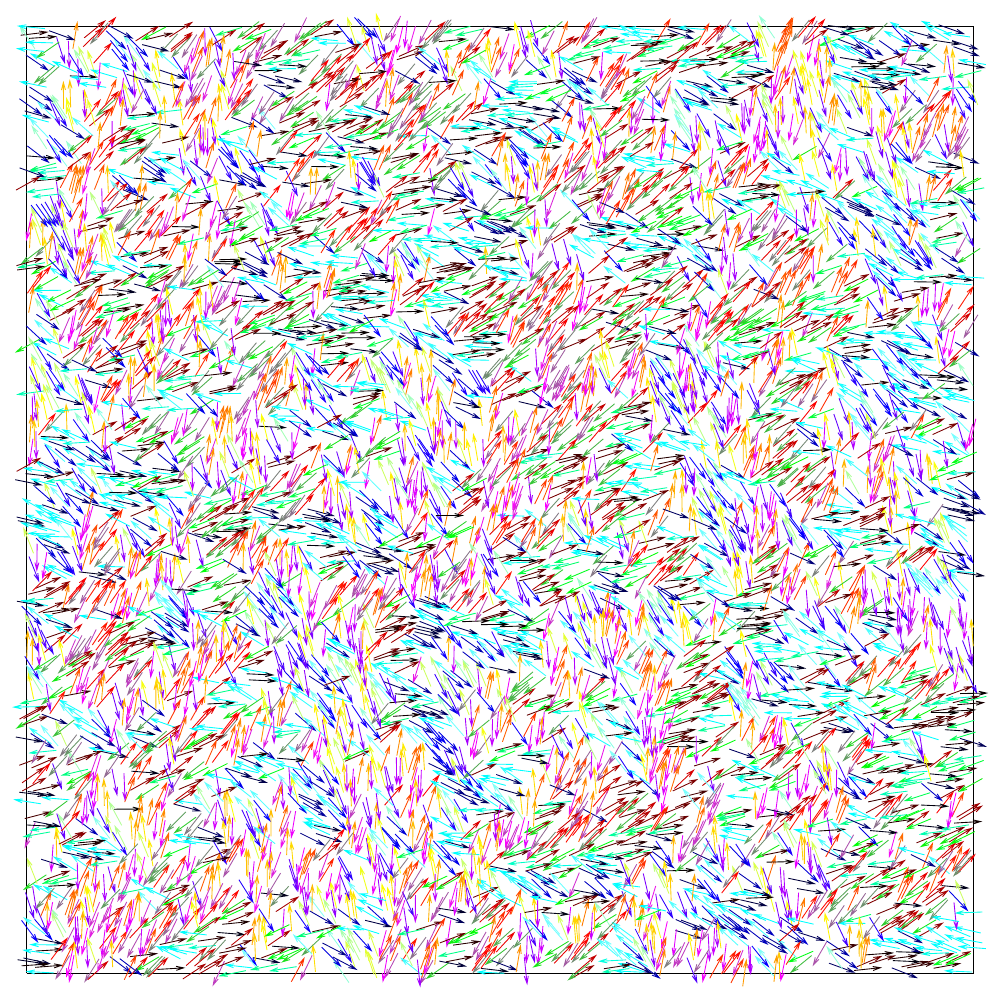}}
  \subfigure[~$\Pe=20,\, \rho\,L_\rod^2=5.1$]{\label{snapshot_c}\includegraphics[totalheight=0.5\columnwidth]{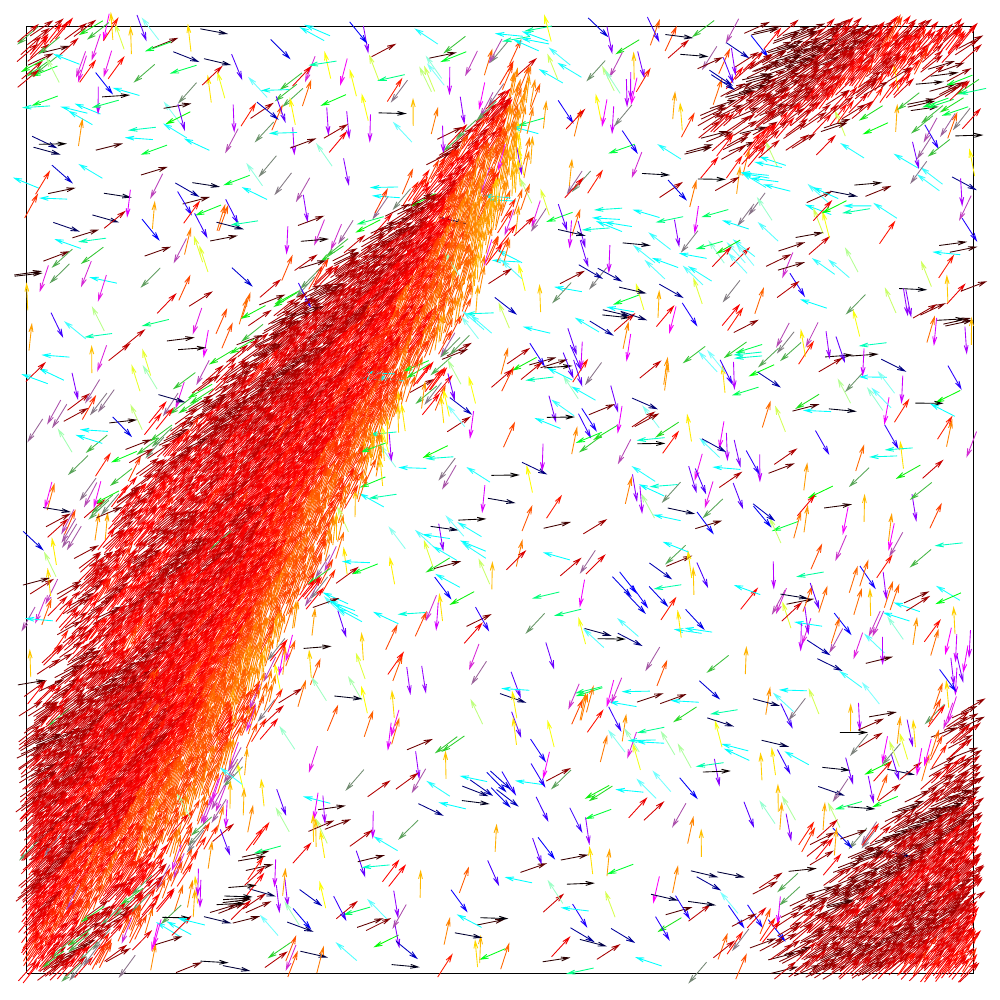}}
  \subfigure[~$\Pe=100,\, \rho\,L_\rod^2=5.1$]{\label{snapshot_d}\includegraphics[totalheight=0.5\columnwidth]{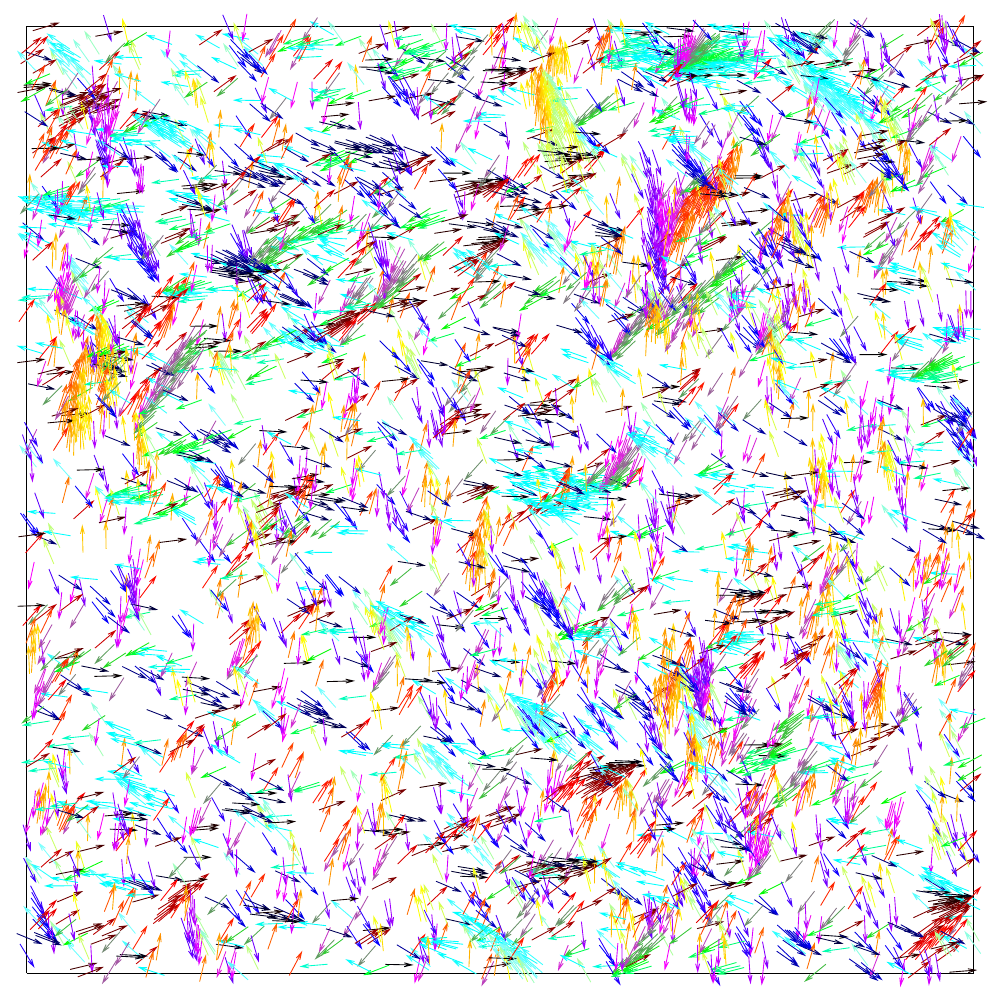}}\\
  \subfigure[~$\Pe=25,\, \rho\,L_\rod^2=10.2$]{\label{snapshot_e}\includegraphics[totalheight=0.5\columnwidth]{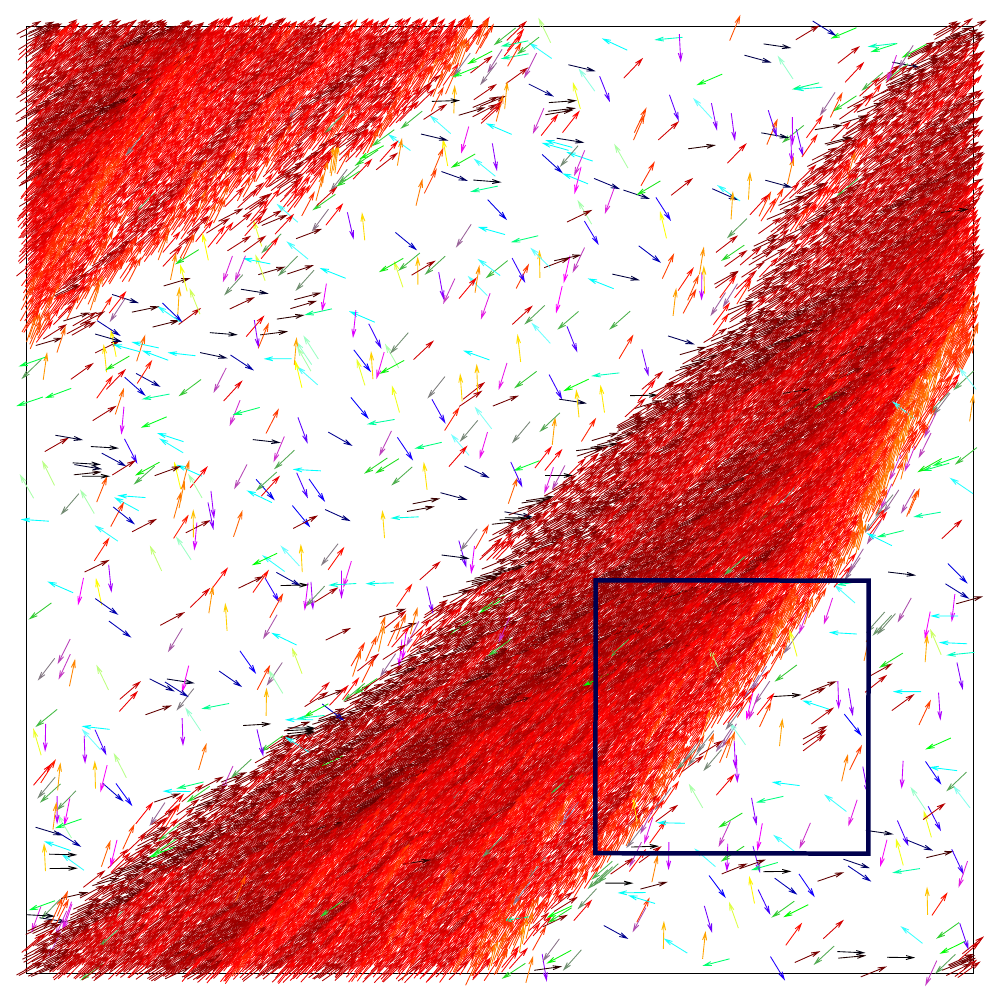}}
  \subfigure[~close-up of (e)]{\label{snapshot_f}\includegraphics[totalheight=0.5\columnwidth]{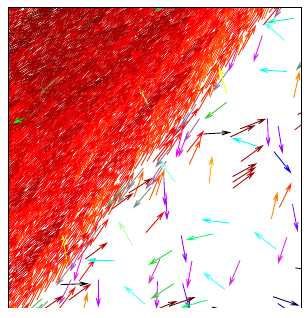}}
  \subfigure[~$\Pe=75,\, \rho\,L_\rod^2=25.5$]{\label{snapshot_g}\includegraphics[totalheight=0.5\columnwidth]{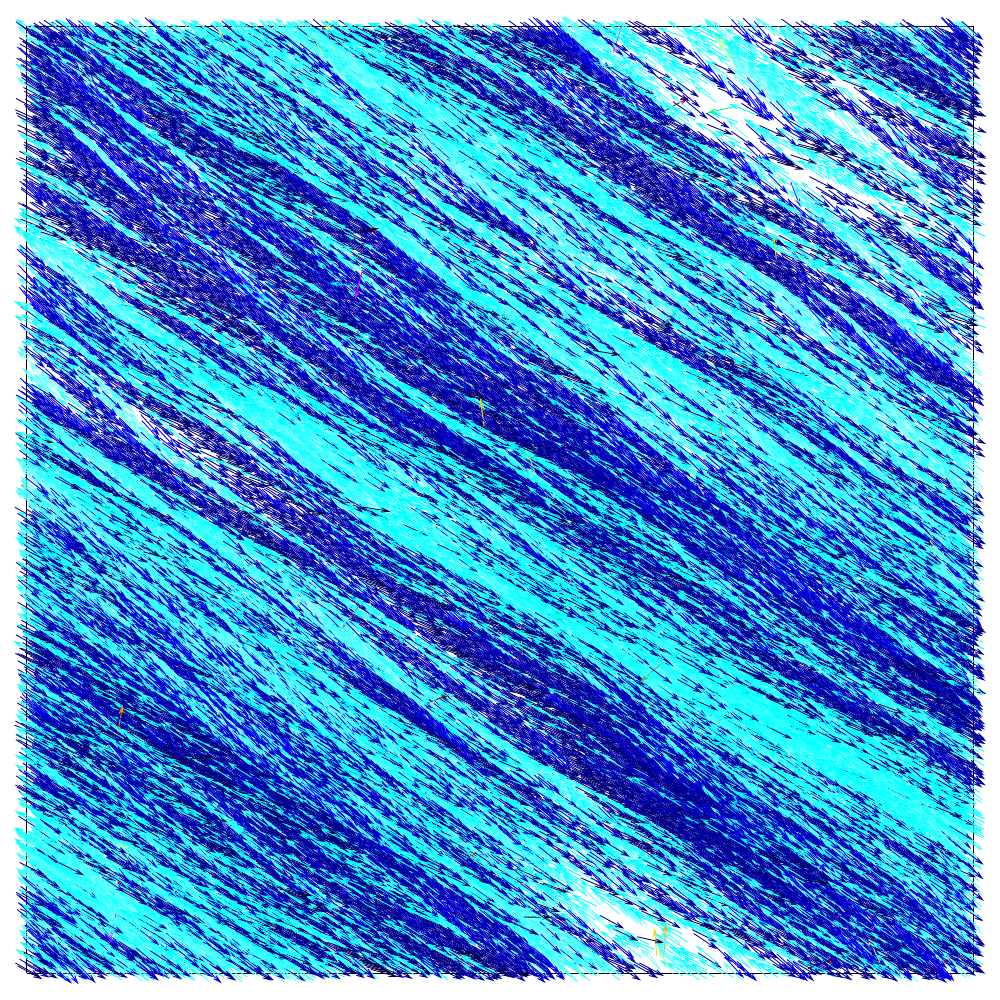}}
  \subfigure[~color coding guide]{\label{colorwheel}\includegraphics[totalheight=0.5\columnwidth]{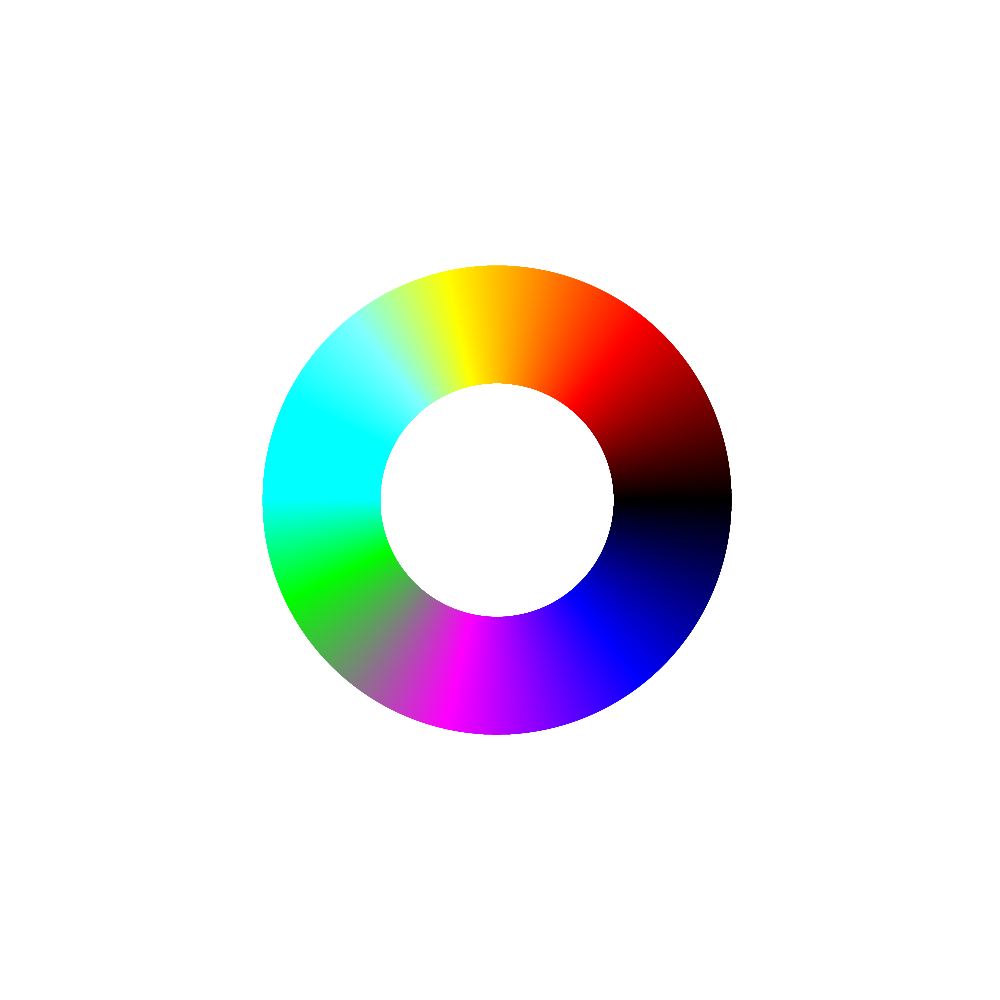}}
\caption{(Color online) Snapshots of self-propelled rod systems simulated using Brownian dynamics simulations. Each rod is colored based on its orientation. (a) Nematic state at high scaled density $\rho\,L_\rod^2$ and zero P\'eclet number $\Pe$, (b) isotropic state at low $\rho$ and zero $\Pe$, (c) and (e) giant clusters at medium $\Pe$, (f) close-up of a boundary of a giant cluster, (d) cluster break-up at high $\Pe$, (g) laning phase at high $\rho$, and (h) color coding for rod orientation. For movies, see the Supplemental Material \cite{Abkenar13suppl}.}
\label{fig:snapshots}
\end{figure*}

\section{Model and Simulation Technique}
\label{sec:model}
\begin{figure}[]
\centering
  \includegraphics[trim= 0.0cm 0.0cm 0.0cm 0cm, clip=true, totalheight=0.5\columnwidth]{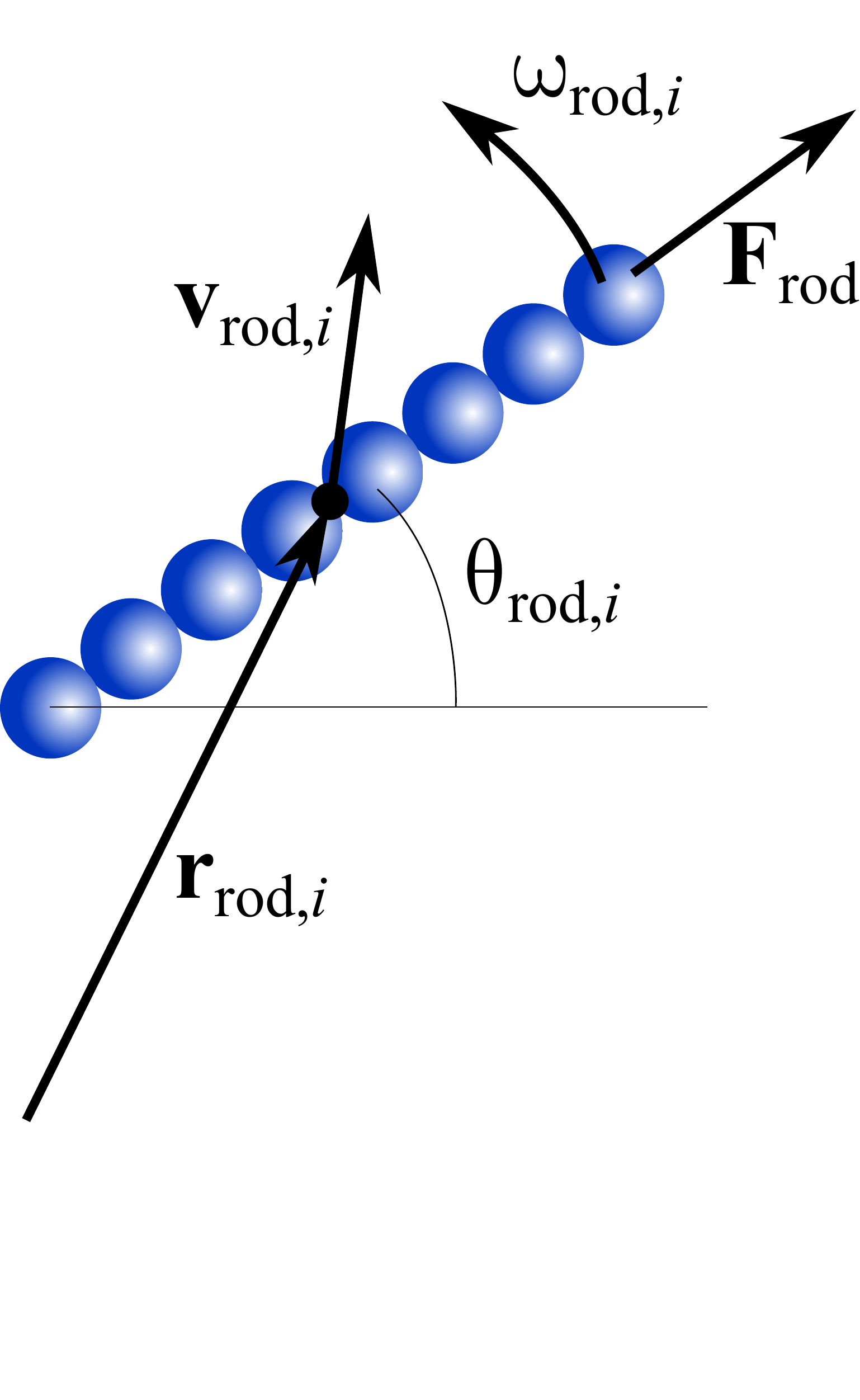}
  \includegraphics[trim= 0.0cm 0.0cm 0.0cm 0cm, clip=true, totalheight=0.56\columnwidth]{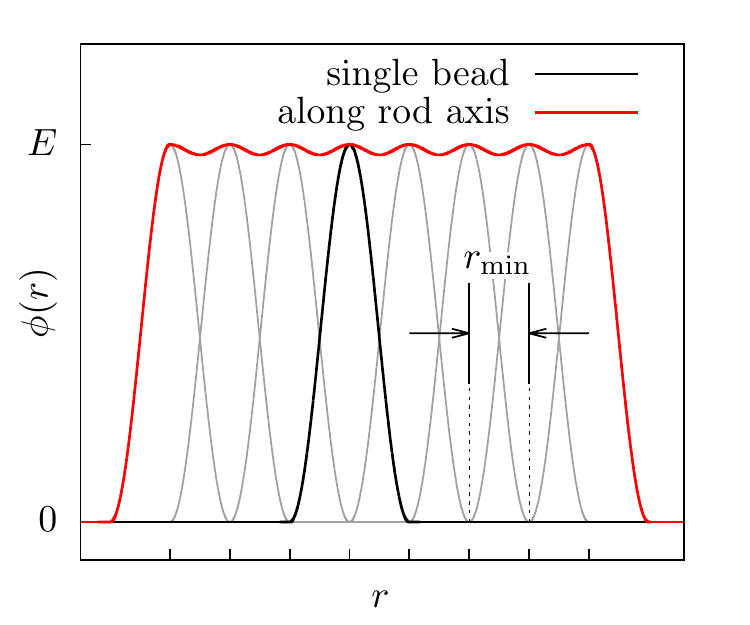}
\caption{(Color online) Left: schematic representation of the model of a self-propelled rod and coordinates used in two dimensions. The rod is discretized into $n_\mathrm{b}$ beads to calculate the rod-rod interaction. Right: potential profile of a rod along its long axis. Tics on the horizontal axis show the position of beads, separated from each other by $r_\mathrm{min}$. In our simulations, we use $n_\mathrm{b}=18$. }\label{fig:RodModel}
\end{figure}

We simulate rods with and without an intrinsic propulsion force.
Our systems consist of $N_\rod$ rods in a two-dimensional box of size $L_x\times L_y$ with periodic boundary conditions; see Fig.~\ref{fig:snapshots}. We use Brownian dynamics simulations for active systems and Monte Carlo simulations for passive systems. 
The rods are characterized by their center-of-mass positions $\mathbf{r}_{\rod,i}$, their orientation angles $\theta_{\rod,i}$ with respect to the $x$ axis, their center-of-mass velocities $\mathbf{v}_{\rod,i}$, and their angular velocities $\omega_{\rod,i}$; see Fig.~\ref{fig:RodModel}.
To calculate energy, force, and torque due to rod-rod interactions, we discretize each rod into $n_\mathrm{b}$ beads, separated from each other by a distance of $L_\rod/n_\mathrm{b}$.
Beads from different rods interact by a separation-shifted Lennard-Jones potential \cite{Fisher1966},
\begin{equation} 
\phi(r) = \left\{ 
\begin{array}{ll} 
4\epsilon\left[(\alpha^2+r^2)^{-6}-(\alpha^2+r^2)^{-3}\right]+\phi_0, &  r<r_\mathrm{min} \\ 
0, & r\ge r_\mathrm{min}\,,
\end{array} \right.
\label{eq:sslj}\end{equation}
where $r$ is the distance between two beads and $\epsilon$ gives the
interaction energy. The potential is shifted by $\phi_0$ to avoid a
discontinuity at $r=r_\mathrm{min}$. The parameter $\alpha$ characterizes
the capping of the potential. For $\alpha\neq 0$, $\phi$ does not diverge
at $r=0$, hence allowing bead-bead overlap; for $\alpha=0$, $\phi(r)$ becomes
the truncated Lennard-Jones potential.

$E=\phi(0)-\phi(r_\mathrm{min})$ is the energy for two beads that completely
overlap and is used as independent parameter in our simulations. Setting $E$ to any value will dictate $\epsilon=\alpha^{12}E/(\alpha^{12}-4\alpha^{6}+4)$. 
The constant $\alpha=(2^{1/3}-r_\mathrm{min}^2)^{1/2}$ is calculated by forcing $\phi(r)$ to be zero at $r=r_\mathrm{min}$. 
Considering the weak repulsion between rods, we define $r=r_\mathrm{min}/2$ as the effective radius for each bead, which results in the effective rod thickness $r_\mathrm{min}$ and the rod aspect ratio $L_\rod/r_\mathrm{min}$.
The number of beads $n_\mathrm{b}$ used for discretization is chosen such that the rod has a relatively smooth potential profile, so that no interlocking occurs when rods slide along each other; see Fig.~\ref{fig:RodModel}.

For the Brownian dynamics simulations, we decompose the rod velocity into parallel and perpendicular components with respect to its axis, $\mathbf{v}_{\rod,i}=\mathbf{v}_{\rod,i,\Vert}+\mathbf{v}_{\rod,i,\perp}$.
In each simulation step, the velocities are calculated using
\begin{eqnarray}
\mathbf{v}_{\rod,i,\Vert}(t)&=&\frac{1}{\gamma_\Vert}\left(\sum_{j\neq i}^{N_\rod}\mathbf{F}_{ij,\Vert}+\xi_\Vert\mathbf{e}_\Vert+F_{\rod}\mathbf{e}_\Vert\right)\,,\\
\mathbf{v}_{\rod,i,\perp}(t)&=&\frac{1}{\gamma_\perp}\left(\sum_{j\neq i}^{N_\rod}\mathbf{F}_{ij,\perp}+\xi_\perp\mathbf{e}_\perp\right)\,,
\end{eqnarray}
and
\begin{eqnarray}
\omega_{\rod,i}(t)&=&\frac{1}{\gamma_r}\left(\sum_{j\neq i}^{N_\rod}M_{ij}+\xi_r\right)\,,
\end{eqnarray}
where $\mathbf{e}_\Vert$ and $\mathbf{e}_\perp$ are unit vectors parallel and perpendicular to the rod axis, respectively.
$F_{\rod}$ is the propulsion force for each rod. The friction coefficients are given by $\gamma_\Vert=\gamma_0 L_\rod$, $\gamma_\perp=2\gamma_\Vert$, and $\gamma_r=\gamma_\Vert L_\rod^2/6$, where $L_\rod$ is the rod length \cite{Yang2010}. The random values $\xi_\Vert$, $\xi_\perp$, and $\xi_r$ for the forces in parallel and perpendicular direction and for the torque are drawn from Gaussian distributions with variances $\sigma_\rod^2L_\rod$, $2\sigma_\rod^2L_\rod$, and $\sigma_\rod^2L_\rod^3/12$, respectively.
We employ thermal noise, thus the variances are calculated using $\sigma_\rod^2=2\kBT/\gamma_0 \Delta t$ \footnote{In biological and synthetic self-propelled systems, the noise arises from the environmental noise, for example, from density fluctuations of signaling molecules for chemotactic swimmers or from motor activity.
In this case, the noise is not proportional to $k_\mathrm{B}T$ and also the coupling between translational and rotational noise may be different.}.
Finally, $\mathbf{F}_{ij}$ and $M_{ij}$ are the force and torque from rod $j$ to rod $i$, calculated using Eq.~(\ref{eq:sslj}).
Hydrodynamic interactions between the rods are largely screened because of the
nearby wall and the high rod density \cite{Berke2008,Elgeti2009,Elgeti2010,Elgeti1302}, and hence are neglected in our simulations.

We study systems with approximately $10\,000$ rods at scaled number densities ranging from $\rho\,L_\rod^2=2.5$ to $10$, where the number density of rods is defined as $\rho=N_\rod/L_xL_y$. 
We measure lengths in units of rod length $L_\rod$,
energies in units of $\kBT$,
and times in units of the orientational diffusion time for a single rod, $\tau_0=1/D_r=\gamma_0 L_\rod^3/6\,\kBT$.
The system size is $L_x=L_y=36\,L_\rod$, 
the cutoff $r_\mathrm{min}=L_\rod/n_b=0.056\,L_\rod$, 
the rod aspect ratio $L_\rod/r_\mathrm{min}=18$, 
the time interval $\Delta t=1.65\times 10^{-4}\,\tau_0$, 
and unless mentioned otherwise, $E=1.5\,\kBT$. 

There are three different energy scales in our system; the thermal energy $\kBT$, the propulsion strength $F_\rod L_\rod$, and the energy barrier $E$. Therefore, there are two dimensionless ratios that characterize the importance of the different contributions:  
the P\'eclet number, defined as \footnote{The P\'eclet number can be alternatively defined as $\Pe_r=v_0/D_rL_\rod$ with the rotational diffusion constant $D_r$. In such case, $\Pe=6\,\Pe_r$.}
\begin{equation}
\Pe=\frac{L_\rod v_0}{D_\Vert} = \frac{L_\rod F_\rod}{\kBT}\,,
\label{eq:peclet}
\end{equation}
which is the ratio of propulsion strength to noise, and the \textit{penetrability} coefficient, $Q$, defined as 
\begin{equation}
Q=\frac{L_\rod F_\rod}{E}\,,
\label{eq:q}
\end{equation}
which is the ratio of propulsion strength to energy barrier. $D_\Vert=\kBT/\gamma_\Vert$ is  the diffusion coefficient parallel to the rod orientation.

We simulate rods with P\'eclet $(\Pe)$ numbers in the range $0\le \Pe < 200$ and penetrabilities in the range $0\le Q < 200$. We change $\Pe$ by changing $F_{\rod}$ for fixed $\sigma^2_\rod$ and $\Delta t$, i.~e., for fixed temperature. We change $Q$ by changing both $F_\rod$ and $E$.

\section{Isotropic-nematic transition for passive systems}
\label{sec:passiverods}
\begin{figure}
\centering
  \includegraphics[trim= 0.0cm 0.0cm 0.0cm 0cm, clip=true, totalheight=0.71\columnwidth]{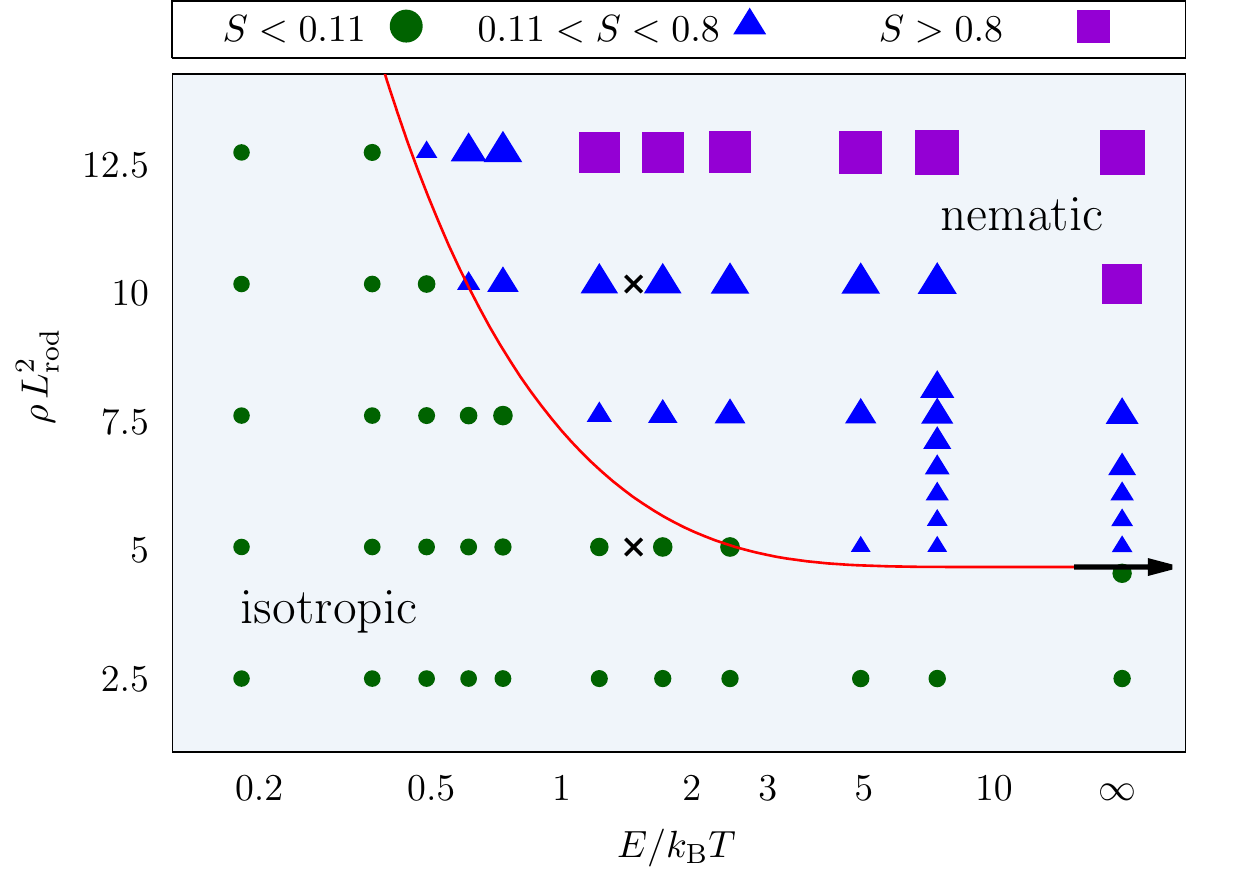}
  \caption{(Color online) Phase diagram for passive rod systems with different densities ($\rho$) and energy barriers ($E$). In addition to color/symbol coding, the size of each triangle is proportional to the nematic order parameter $S$ [Eq.~(\ref{eq:nematicOP})]. Bottom left: isotropic phase with $S<0.11$; top right: nematic phase with $S>0.8$; middle: nematic phase with $0.11<S<0.8$. The black arrow indicates Onsager's isotropic-nematic transition density, $\rho_c\,L_\rod^2 = 3 \pi/2$. The red (gray) line is given by Eq.~(\ref{eq:IsoNem}). Crosses $(\mathbf{\times})$ mark the parameters that have been used for the Brownian dynamics simulations in Figs.~\ref{snapshot_a} and \ref{snapshot_b}.}
\label{fig:IsoNemaPhaseDiag}
\end{figure}

Suspensions of passive rodlike particles in thermal equilibrium 
are isotropic for low densities and nematic for high
densities \cite{Kayser1978}. 
For $E\rightarrow \infty$ and $L_\rod/r_{\rm min}\gg 1$, the transition density 
$\rho_c\,L_\rod^2 = 3 \pi/2$ 
has been predicted using Onsager's theory for infinitely thin hard rods \cite{Kayser1978,Onsager1944}.  
For our systems with the capped potential given by Eq.~(\ref{eq:sslj}), not only
the aspect ratio of the rod but also the energy barrier $E$ affects the
density for the isotropic-nematic transition.
As $E$ becomes smaller, the tendency for rods to align becomes weaker because overlaps occur more frequently.
For $E=0$, the rods do not interact mutually and thus are in the isotropic phase for all densities.

We performed Brownian dynamics simulations for systems with $\Pe=0$ at various densities.
At low densities, $2.5\le\rho\,L_\rod^2\le 5.1$, the systems are in an isotropic state as shown in Fig.~\ref{snapshot_b}. 
For high densities, $7.7\le\rho\,L_\rod^2\le 10.2$, nematic states are found that are composed of large interlocked groups of rods with similar orientations; see Fig.~\ref{snapshot_a}. 
Because the simulation of passive rods with Brownian dynamics is computationally very expensive,
we used Monte Carlo simulations to systematically study the state of the system for several values of $\rho$ and $E$. We characterize the state using the nematic order parameter \cite{Kraikivski2006},
\begin{equation}
S = \left<\sum_{i\neq j}^{N} \frac{1}{N(N-1)} \cos[2(\theta_i-\theta_j)]\right>\,,
\label{eq:nematicOP}\end{equation}
where the average is over cells of side length $4.5\,L_\rod$.
$S=0$ and $S=1$ correspond to perfectly isotropic and nematic states, respectively.

Figure~\ref{fig:IsoNemaPhaseDiag} shows a phase diagram of the system with varying density and energy barrier.
According to analytical theory \cite{Kayser1978}, for $E=\infty$ the transition from the isotropic to the nematic state occurs at $\rho_c\,L_\rod^2=3\pi/2$,
as indicated by the black arrow in Fig.~\ref{fig:IsoNemaPhaseDiag}. This density corresponds to $S=0.11$, which we thus define as threshold value to calculate the transition density for finite values of the energy barrier; see Appendix \ref{app:passivetransitions}.
We have also calculated the density for the isotropic-nematic transition,
\begin{equation}
\rho_c\,L_\rod^2=\frac{3\pi}{2} \frac{1}{[1-\exp(-E/\kBT)]}\,,
\label{eq:IsoNem}
\end{equation}
by generalizing Onsager's approach for finite energy barriers, as described in Appendix~\ref{app:passivetransitions}.
We find very good agreement between 
the analytical theory shown by the red (gray) line in Fig.~\ref{fig:IsoNemaPhaseDiag}
and our Monte Carlo simulations. 
The phase diagram is also consistent with our Brownian dynamics simulations for $\Pe=0$ and $E=1.5\,\kBT$; see snapshots in Figs.~\ref{snapshot_a} and \ref{snapshot_b}.

\section{Crossing probability for rod-rod collisions}
\label{sec:crossing}
\begin{figure}
\centering
  \includegraphics[trim= 0.0cm 0.0cm 0.0cm 0cm, clip=true, totalheight=0.67\columnwidth]{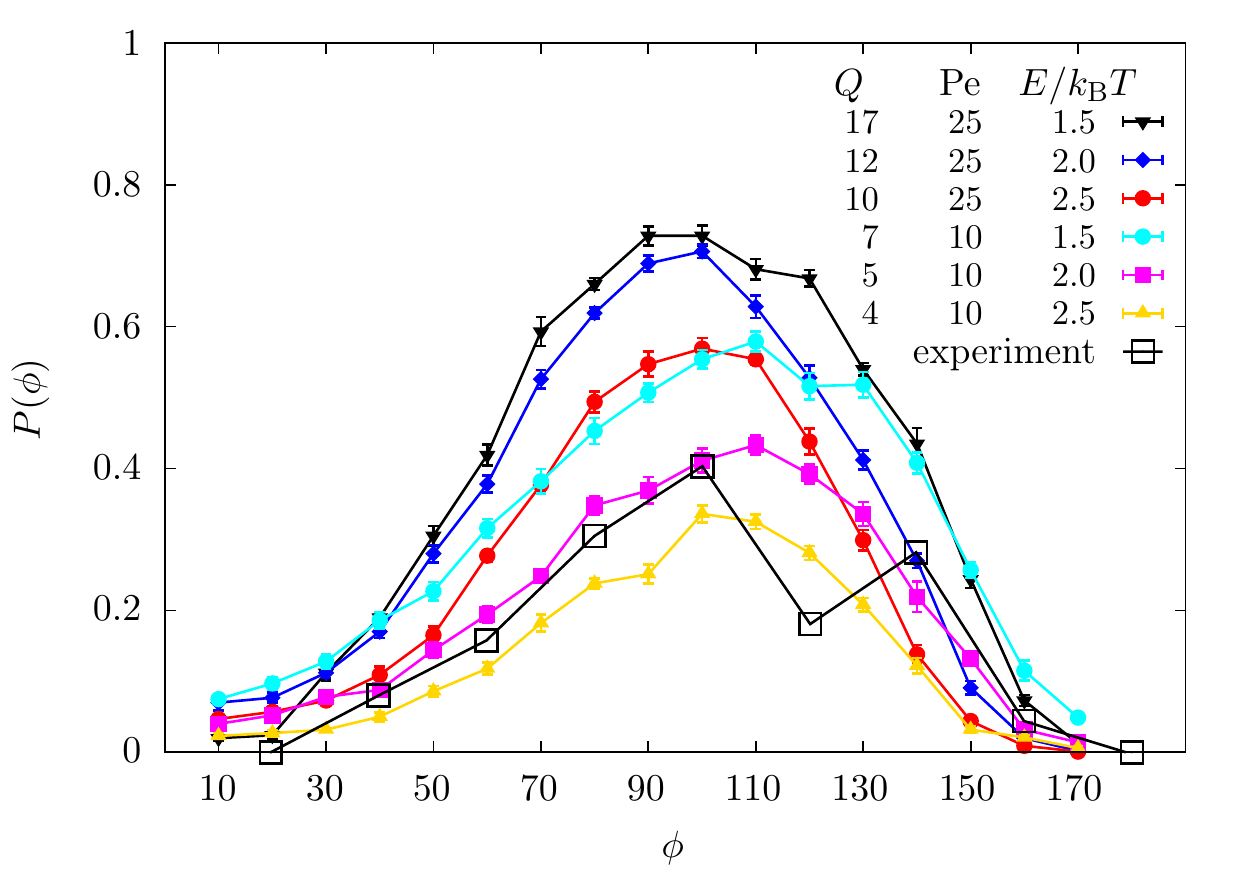}\\
  \includegraphics[trim= 0.0cm 1.5cm 0.0cm 0cm, clip=true, totalheight=0.30\columnwidth]{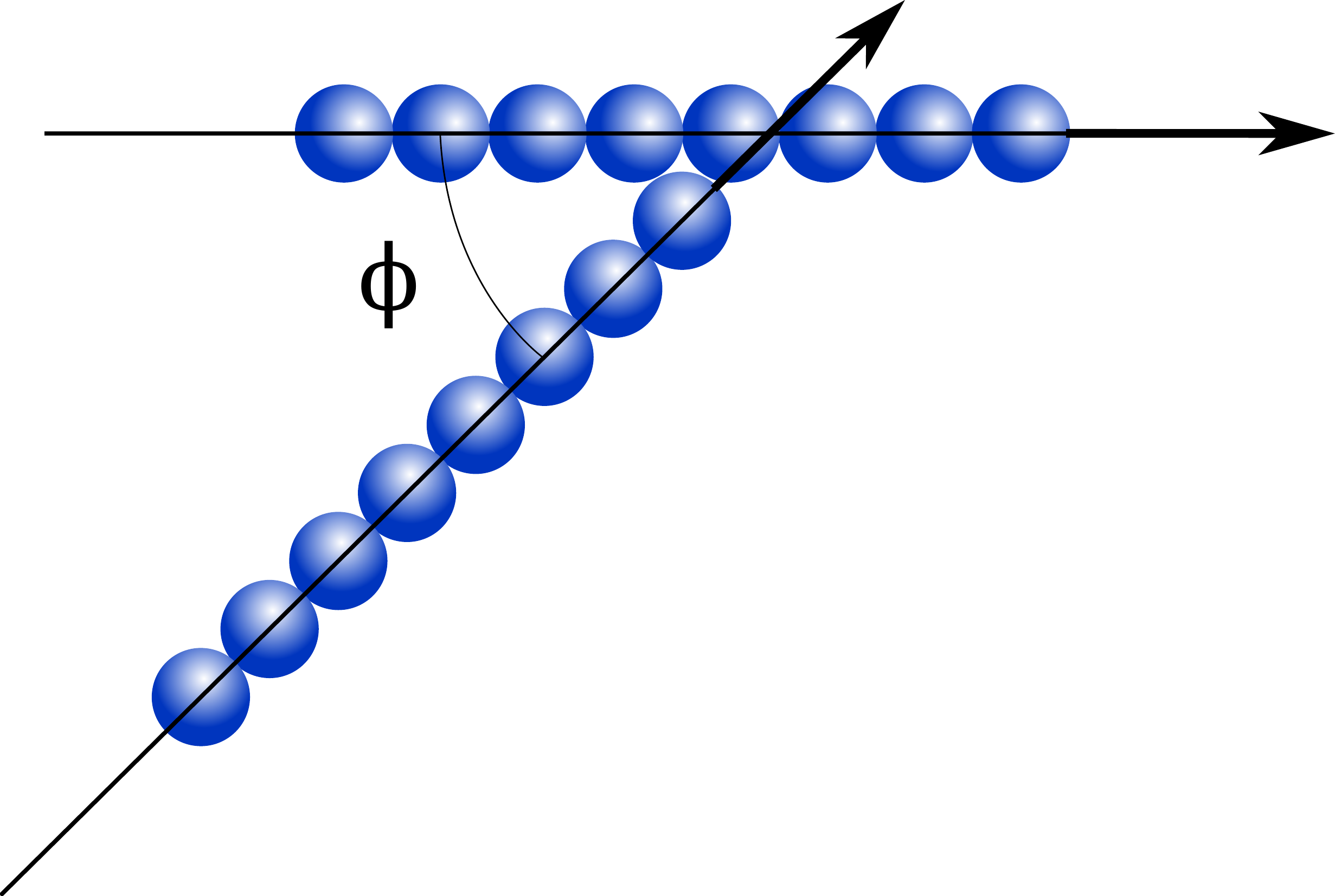}
\caption{(Color online) Crossing probability for two rods as a function of their crossing angle $\phi$ (as defined in the schematic). For each angle, $1000$ simulations have been performed. The simulations are divided into $10$ groups and the error bars are calculated as the standard deviation of the mean $(\sigma_m)$ for these groups. The experimental data are taken from Ref.~\cite{Sumino2012}.}
\label{fig:CrossingRates}
\end{figure}
To find the probability of crossing events $P(\phi)$, we performed simulations for two rods that initially touch each other in a tip-center arrangement with crossing angle $\phi$; see Fig.~\ref{fig:CrossingRates}. We measure $P(\phi)$ for several penetrabilities and P\'eclet numbers using Brownian dynamics simulations. 
We count a crossing event when two rods intersect significantly, i.~e., such that the intersection point is at least $0.2L_\rod$ away from the ends of each rod.
We thus do not count events when one rod only ``touches'' the other rod, which frequently happens due to the weak repulsion between the rods. 

As shown in Fig.~\ref{fig:CrossingRates}, $P(\phi)$ is low near $\phi\simeq 0^{\circ}$ and $\phi\simeq 180^{\circ}$ and has a peak near $\phi\simeq 90^{\circ}$. There is a small asymmetry in the peak with an enhancement for directions $\phi>90^{\circ}$, which may be attributed to the increased relative velocity between two rods for $\phi>90^{\circ}$ and the fact that the rods are not perfectly smooth. 
Comparison between $P(\phi)$ for different penetrabilities shows that an increased $Q$ generally increases the probability for rod crossing.
In addition, for small $\Pe$, noise also plays an important role to enhance rod crossing.
For example, the curves for $Q=10$ and $Q=7$ in Fig.~\ref{fig:CrossingRates} have approximately the same height, and this could be explained by the fact that the effect of noise is higher for the case $Q=7$ that has a smaller $\Pe$.

The results are qualitatively similar to the crossing probability measured in experiments with
microtubules propelled on surfaces. In Fig.~3(d) in Ref.~\cite{Sumino2012}, the maximum crossing probability for two microtubules in a motility assay is $40 \, \%$ and corresponds to $Q=5$ and $\Pe=10$ in our simulations. 
However, the same crossing probability may be achieved by reducing $Q$ and increasing $\Pe$ at the same time.

\section{Cluster formation for active systems}
\label{sec:clusterformation}
\begin{figure}
\centering
  \includegraphics[trim= 0.0cm 0.0cm 0.0cm 0cm, clip=true, totalheight=0.71\columnwidth]{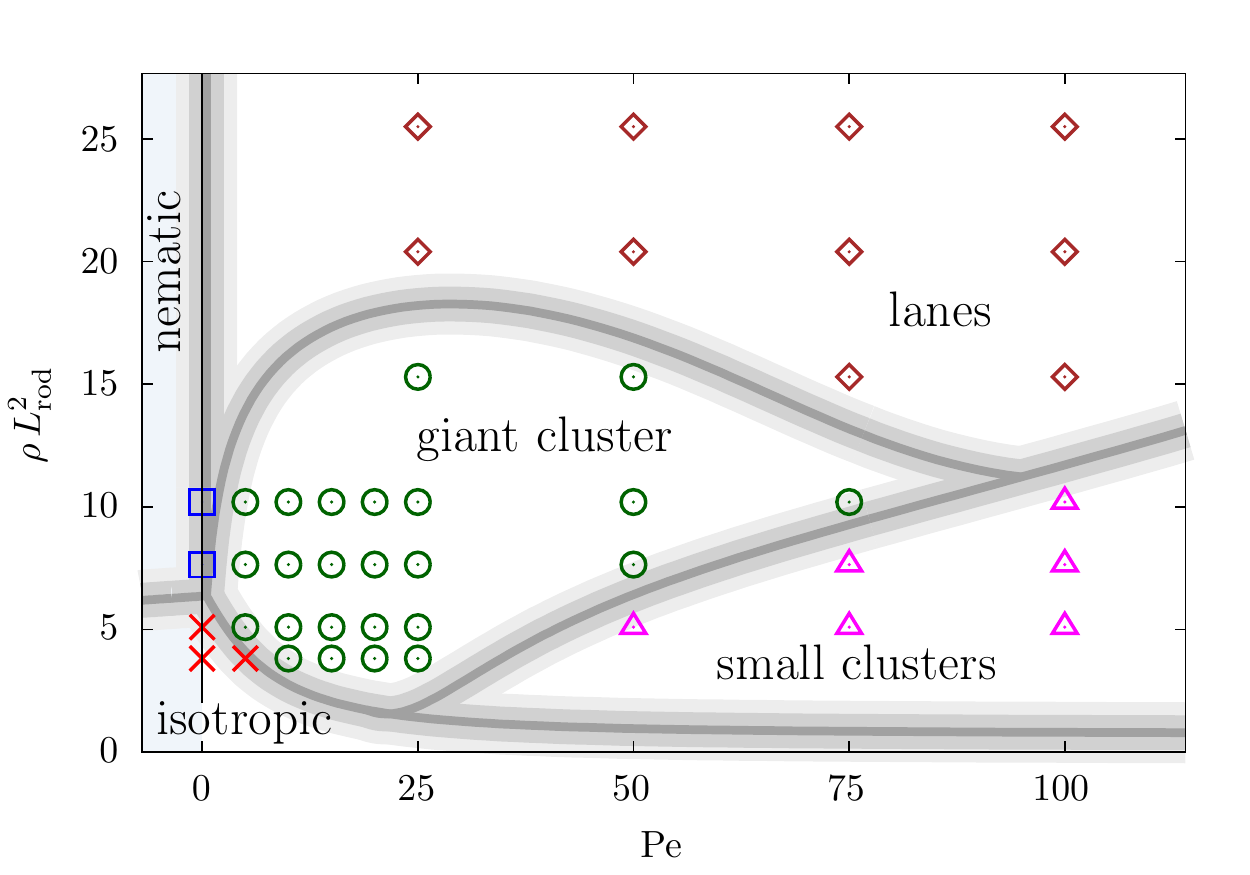}
  \caption{(Color online) Phase diagram for self-propelled rods with different densities ($\rho$) and P\'eclet numbers ($\Pe$). The energy barrier is $E=1.5\,\kBT$; the gray lines are guides to the eye. Note that the region $\Pe<0$ has no physical meaning and that the nematic state is found for passive rods with $\Pe=0$.}
\label{fig:ActivePhaseDiag}
\end{figure}
To characterize the collective behavior, we have performed simulations with large numbers of rods.
After initiating the rods with random positions and orientations in the two-dimensional (2D) plane, the rods move by their propulsion force and are affected by interactions with other rods and thermal noise. 
Snapshots of the system are shown in Fig.~\ref{fig:snapshots}. More snapshots and movies can be found in the Supplemental Material \cite{Abkenar13suppl}. 
A phase diagram of self-propelled rods with varying density and P\'eclet number is shown in Fig.~\ref{fig:ActivePhaseDiag}.

For $1\le \Pe\lesssim 80$, we find giant clusters that span the entire simulation box and form as a result of the alignment interaction due to the rod-rod repulsion, as explained qualitatively in Refs.~\cite{Baskaran2012,Ginelli2010}; see Figs.~\ref{snapshot_c} and \ref{snapshot_e}. 
At the cluster perimeter, the clusters steadily lose rods due to the rotational diffusion and at the same time acquire new rods that collide and align. 
The clusters are polar and almost all rods within a giant cluster move in the same direction.
However, we expect that the system is essentially in an isotropic phase, and that for a sufficiently large system size the clusters can randomly change direction. 
The polar order of our giant clusters which span the simulation box is due to symmetry-breaking collisions because of the roughness of the rods.
In the early stage of the formation of giant clusters, some of the eventually polar clusters are composed of streams of rods that move in opposite directions.

Upon further increase of $\Pe$ the clusters start to break; see Fig.~\ref{snapshot_d}. Smaller clusters are observed until they become as small as about five rods per cluster for $\Pe\gtrsim 100$. 
For very high densities, $15.1\le\rho\,L_\rod^2\le 25.5$, when the dense region spans the entire simulation box, we find a laning phase that is composed of streams of rods that move in opposite directions; see Fig.~\ref{snapshot_g}. 
The laning phase is nematic, similar to the nematic lanes that have been observed for the Vicsek model in simulations \cite{Ginelli2010} and analytical calculations \cite{Peshkov2010}.

Our phase diagram in Fig.~\ref{fig:ActivePhaseDiag} may be compared with the phase diagram in Ref.~\cite{Wensink2012} for self-propelled rods that interact segment-wise via a Yukawa potential. 
Since our model incorporates noise and has a capped repulsive interaction potential, we can only compare both models in the medium $\Pe$ regime, where the noise does not dominate $(\Pe\gg 1)$ and where the rods are not completely penetrable $(\Pe \lesssim 75)$.
For aspect ratio $18$ used in our simulation, we see qualitatively similar behavior with increasing density, namely the transition from the isotropic phase to the swarming (clustering) phase and then to the laning phase.

A comparison of our phase diagram in Fig.~\ref{fig:ActivePhaseDiag} with that of Ref.~\cite{Yang2010} shows that we do not observe jammed giant clusters as reported in Ref.~\cite{Yang2010}, because we employ a smoother potential profile along the rod; see Fig.~\ref{fig:RodModel}.

\subsection{Rod densities}
\label{sec:voiddensity}
\begin{figure}
\centering
  \includegraphics[trim= 0.0cm 0.0cm 0.0cm 0cm, clip=true, totalheight=0.68\columnwidth]{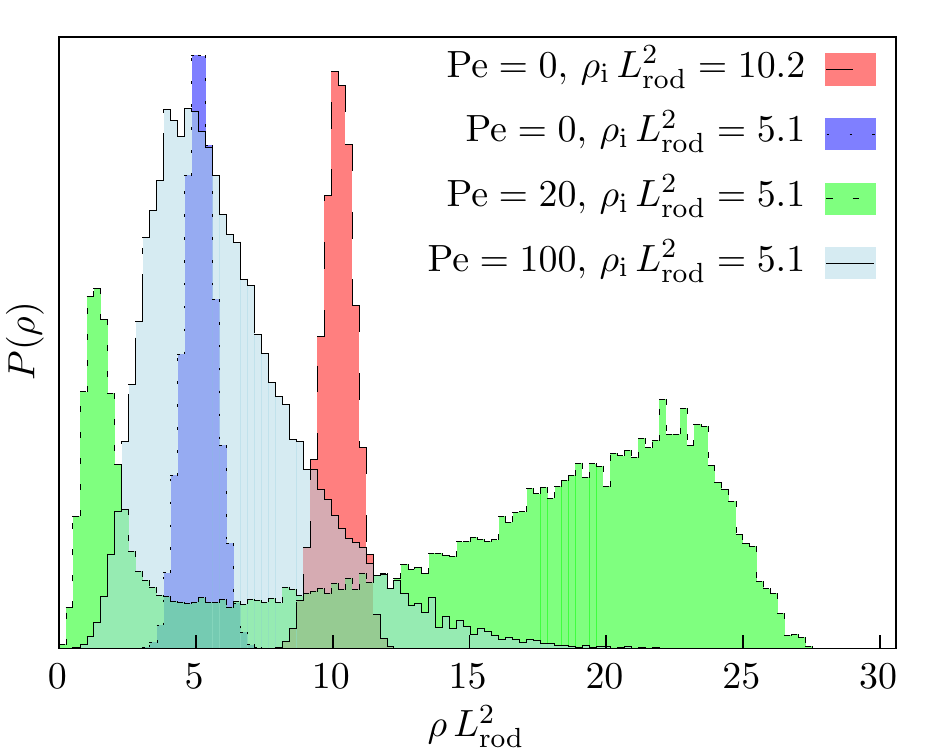}
\caption{(Color online) Density distributions for the systems shown in snapshots of Fig.~\ref{fig:snapshots}. $\rho_\mathrm{i}$ is the average density. The distributions are not normalized and only the position of peaks can be compared.
}
\label{fig:DensityDistSamples}
\end{figure}
\begin{figure}
\centering
  (a)\includegraphics[trim= 0.0cm 0.0cm 0.0cm 0cm, clip=true, totalheight=0.7\columnwidth]{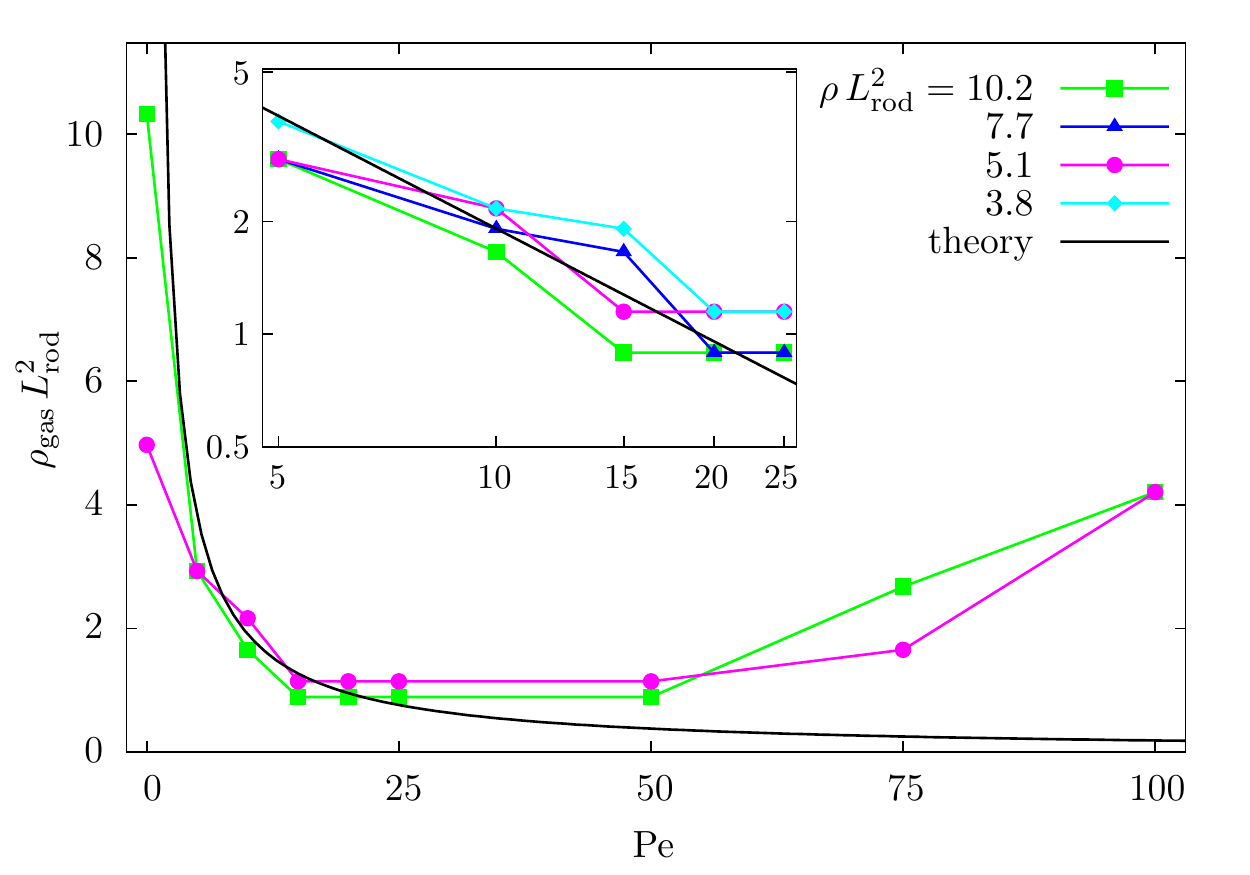} \\
  (b)\includegraphics[trim= 0.0cm 0.0cm 0.0cm 0cm, clip=true, totalheight=0.7\columnwidth]{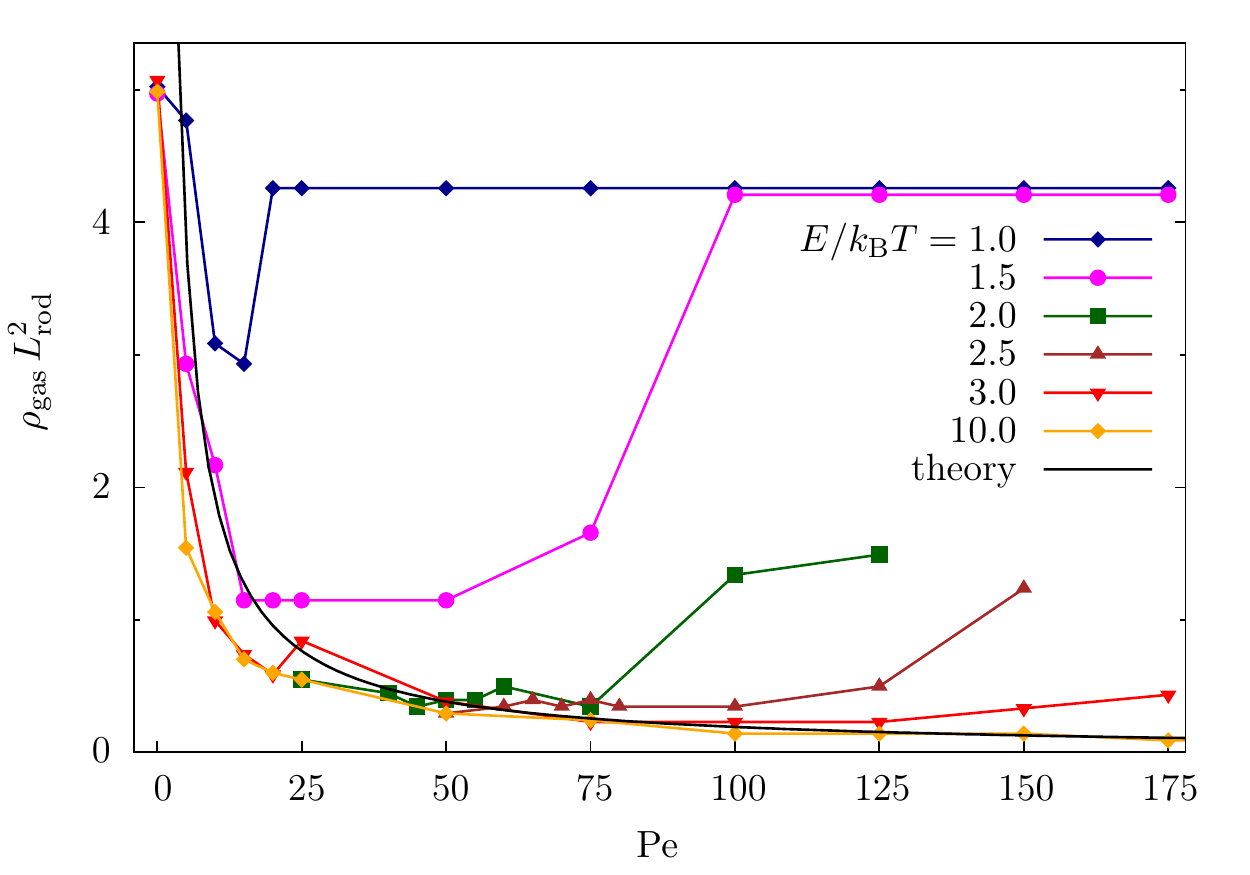}
\caption{(Color online) Gas density as a function of the P\'eclet number for different average rod densities and energy barriers compared with the estimate in Eq.~(\ref{eq:rhovoid}). (a) Gas densities for $E=1.5\, k_BT$ and several rod densities. Inset: double-logarithmic plot of $\rho_\gas$ for $5<\Pe<25$. 
~(b) Gas densities for average rod density $\rho\,L_\rod^2=5.1$ and several energy barriers. The errors are given by the peak width for the density histograms, $\rho\,L_\rod^2\simeq 0.5$.}
\label{fig:rhovoidall}
\end{figure}
We measure densities of rods in cells of side length $2\,L_\rod$ and construct a distribution of monomer densities for each system; see Fig.~\ref{fig:DensityDistSamples}.
For a homogeneous system of rods, the distribution has a single narrow peak at the average density of the system, $\rho_i$.
This can be seen for example in the histograms for $\Pe=0$ that correspond to the systems where no cluster formation is observed; see Figs.~\ref{snapshot_a} and \ref{snapshot_b}.
For systems with self-propelled rods, the density distribution can change from a binomial to a more complicated distribution that shows phase separation between dilute and dense regions of rods. For $\Pe=20$ the distribution has a large peak at low density and a very broad peak at higher densities. 
The noise in the distribution is due to the poor statistics in the intermediate density regime. 
The system consists of a (high-density) cluster in a ``gas'' of rods; the density of this cluster-free region corresponds to the position of the first peak in the density distribution. In the following, we denote the density of this cluster-free region as $\rho_\gas$. 

In Fig.~\ref{fig:rhovoidall},
$\rho_\gas$ is plotted as a function of $\Pe$ for
several values of $\rho_i$ and $E$. 
We define $\rho_\gas$ as the position of the first local maximum in the density distribution, which is at least as high as $80\%$ of the absolute maximum.
We find $\rho_\gas \sim \Pe^{-1}$ for small 
$\Pe$ and an increase of $\rho_\gas$ with increasing $\Pe$ for high $\Pe$. The
gas density is to a large extent independent of the average
rod density of the entire system;see Fig.~\ref{fig:rhovoidall}(a). 
This behavior is analogous
to the \textit{vapor density} for liquid-gas phase coexistence in conventional liquids, where the density of the gas phase only depends on the temperature and is independent of the volume of the liquid phase.

The dependence of $\rho_\gas$ on $\Pe$ and $\rho$ in the low $\Pe$ range can be quantitatively explained by a rate equation~\cite{redner1301}. In the stationary state, the rate of rods joining a cluster equals the rate of rods leaving a cluster.
Assuming an isotropic distribution of rods in the gas, the number of rods joining the cluster from an infinitesimally small box of side length $\ud x$ and $\ud y$ is $\ud^3 N=\rho\,\ud x\,\ud y\,\ud t(1-\cos\theta)/2$, where $\theta=\cos^{-1}(\ud x/(v\ud t))$ is the half angle of a cone inside which rods reach the wall in a given time $\ud t$, and $x$ is the distance to the cluster ``wall.'' Integrating $\ud^3 N$ over $x$ from $0$ to $v\ud t$ gives the attachment rate
\begin{equation}
\Jatt=\frac{\ud^2 N}{\ud t\ud y}=\frac{\rho v}{4}=\frac{\rho D_\Vert \Pe}{4L_\rod}\,,\label{eq:jatt}
\end{equation}
where we have used the definition of P\'eclet number in Eq.~(\ref{eq:peclet}).

The detachment rate $\Jdet$ is determined by the rotational diffusion of the rods;
the typical time a rod needs to diffuse by an angle $\alpha$ is
\begin{equation}
\tau=\frac{\alpha^2}{2D_r}\,.
\end{equation}
Assuming that a complete detachment from the cluster requires $\alpha=\pi/2$ and that rods are placed regularly along the border of a cluster, the detachment rate is found to be
\begin{equation}
\Jdet=\frac{1}{L_\rod\tau}=\frac{8D_r}{\pi^2 L_\rod}\,.\label{eq:jdet}
\end{equation}

By equating $\Jatt$ and $\Jdet$, we find $\rho_\gas$ as a function of $\Pe$,
\begin{equation}
\rho_\gas=\frac{32}{\pi^2}\frac{D_r}{D_\Vert}\frac{1}{\Pe}=\frac{192}{\pi^2L_\rod^2}\frac{1}{\Pe}\,,
\label{eq:rhovoid}
\end{equation}
where we have used $D_r/D_\Vert=\gamma_\Vert/\gamma_r=6/L_\rod^2$. 
Note that the gas density in Eq.~(\ref{eq:rhovoid}) only depends on $L_\rod$ and $\Pe$ and is independent of the average system density $\rho$, which is consistent with the simulation results. 
This implies that the giant cluster grows until the density of the dilute region reaches $\rho_\gas$.

Note that this estimate includes several approximations, in particular using free
rotational diffusion for rods at the border of the cluster and assuming that
complete detachment requires the rods to diffuse by $\alpha = \pi/2$.
As shown in Fig.~\ref{fig:rhovoidall}, the analytical estimate in Eq.~(\ref{eq:rhovoid}) agrees well with the simulation results in the small-$\Pe$ range without any adjustable parameters.
Assuming a two-dimensional gas for the dilute rod phase, we can thus estimate an
effective binding energy per rod for the rods inside the giant cluster,
\begin{eqnarray}
E_b & = & \kBT \ln(\rho_\gas L_\rod r_{\rm min}) \nonumber \\
  & = & \kBT \left[\ln\left(192/\pi^2\right)-\ln(\Pe\, L_\rod/r_{\rm min})\right] \,,
\label{eq:bindingenergy}
\end{eqnarray}
as explained in Appendix \ref{app:bindingenergy}. 
The effective binding strength increases logarithmically with the product of P\'eclet number and the rod aspect ratio. 
For aspect ratio $18$ and $\Pe\approx 25$ used in our simulations, we find effective binding energies of $E_b\approx -0.1\,\kBT$, which are comparable to binding energies for the gas-liquid critical point for colloidal systems \cite{Vliegenthart2000}.

\label{sec:clusterbreakup}
\begin{figure}
\centering
  \includegraphics[trim= 0.0cm 0.0cm 0.0cm 0cm, clip=true, totalheight=0.71\columnwidth]{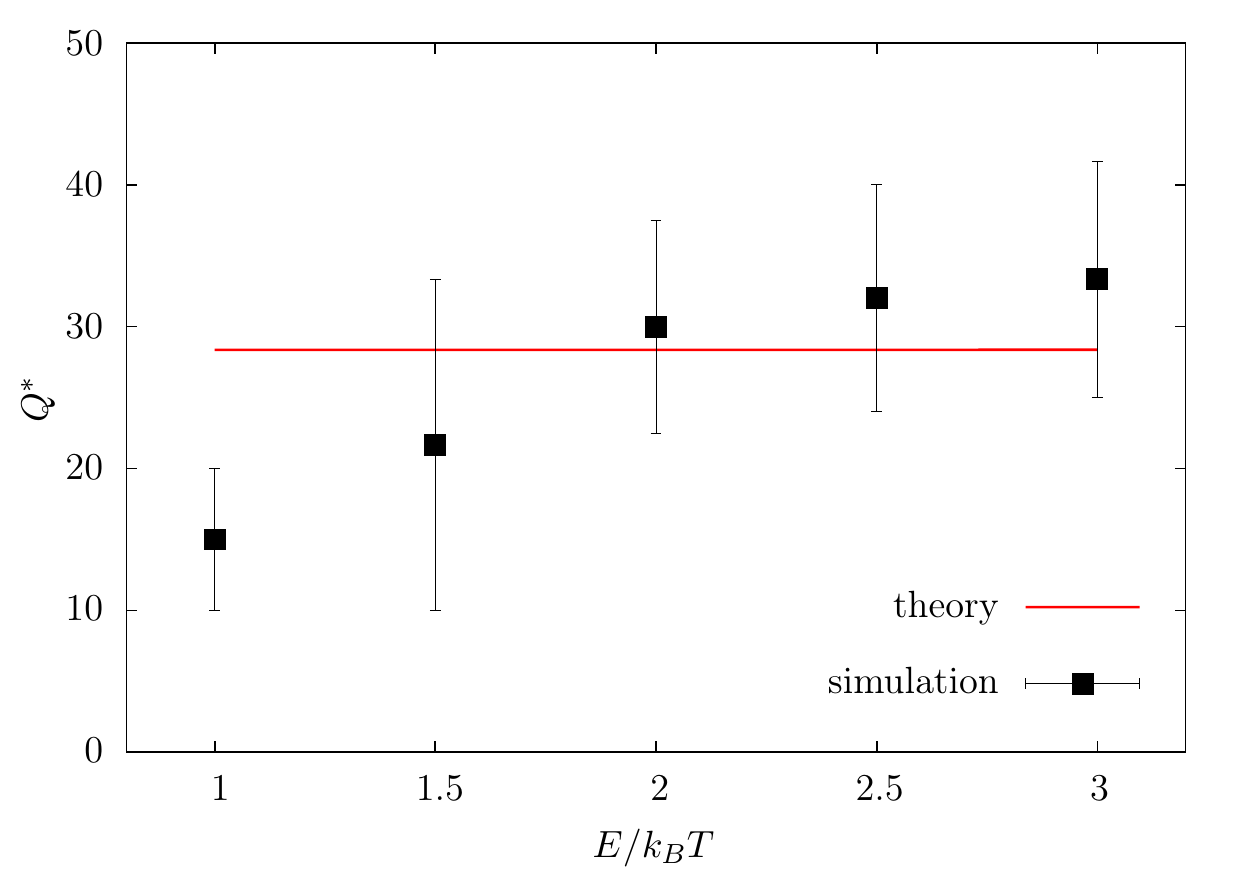}
  \caption{Critical penetrability coefficient $Q^*$ at which clusters start to break vs the energy barrier $E$, compared with the analytical estimate given by Eq.~(\ref{eq:PeBreakUp}). Average rod density is $\rho\,L_\rod^2=5.1$. The points from the simulations are the P\'eclet numbers at which $\rho_\gas$ has a minimum for each energy barrier; compare Fig.~\ref{fig:rhovoidall}.}
\label{fig:PeBreakUp}
\end{figure}
For $E = 1.5 k_B T$, clusters break up when $\Pe \gtrsim 80$, which implies $Q\gtrsim 50$.
We observe that in the regime of cluster break-up, individual rods and
even small clusters can pass through each other. In our simulations $\Pe$
is proportional to the propulsion force, and a high propulsion force thus
facilitates crossing of rods. As a result, fewer rods aggregate in a large cluster and the rod density in the dilute
region $\rho_\gas$ increases; see Fig.~\ref{fig:rhovoidall}. 
Cluster break-up starts when the propulsion force, $F_\rod=Q\, E /L_\rod$, is comparable with the maximum force for bead-bead interaction, $F_\mathrm{int}=\max(-\ud\phi/\ud r)$.
Equating $F_\mathrm{int}$ to $F_\rod$ gives the critical value of the penetrability coefficient for cluster break-up,
\begin{equation}
Q^*=\frac{F_\mathrm{int}L_\rod}{\,E}=28\,,
\label{eq:PeBreakUp}
\end{equation}
where $F_\mathrm{int}=-\ud\phi/\ud r\vert_{r=r_0}$ and $r_0=0.192$ is
found by numerically solving $\ud^2\phi(r)/\ud r^2=0$ for the potential
in Eq.~(\ref{eq:sslj}). In Fig.~\ref{fig:PeBreakUp}, $Q^*$ is plotted
for various energy barriers. Although the angular dependence
for crossing of rods (Fig.~\ref{fig:CrossingRates}) is neglected in the estimate in Eq.~(\ref{eq:PeBreakUp}), we find reasonable agreement with the simulation results without any adjustable parameters. 
However, there is less agreement for small energy barriers, corresponding to small $\Pe$.
The deviations may be accounted for by the noise that for small $\Pe$ is comparable with the propulsion force (but that is not considered in the analytical estimate).

\subsection{Cluster size distributions}
\label{sec:csd}

\begin{figure}
\centering
  \includegraphics[trim= 0.0cm 0.0cm 0.0cm 0cm, clip=true, totalheight=0.71\columnwidth]{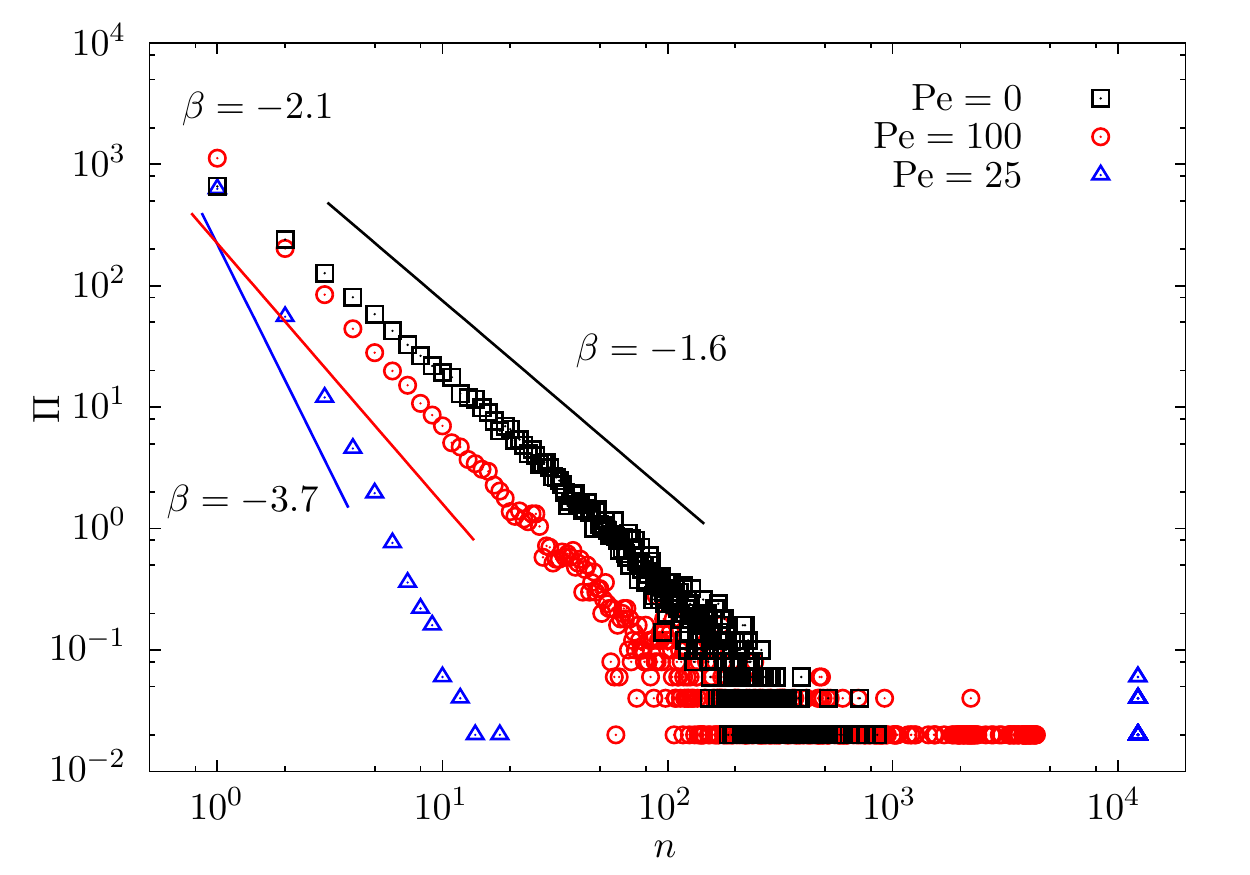}
  \caption{Cluster size distributions $\Pi(n)$ for systems shown in the snapshots of Fig.~\ref{fig:snapshots}. Average rod density is $\rho\,L_\rod^2=10.2$. For small $n$, the distributions can be fit by a power law, $\Pi(n)\propto n^{\beta}$.
The distributions have been averaged over $200$ frames in the last $40\,000$ time steps.
}
\label{fig:sslj178csd}
\end{figure}
\begin{figure}
\centering
  \includegraphics[trim= 0.0cm 0.0cm 0.0cm 0cm, clip=true, totalheight=0.71\columnwidth]{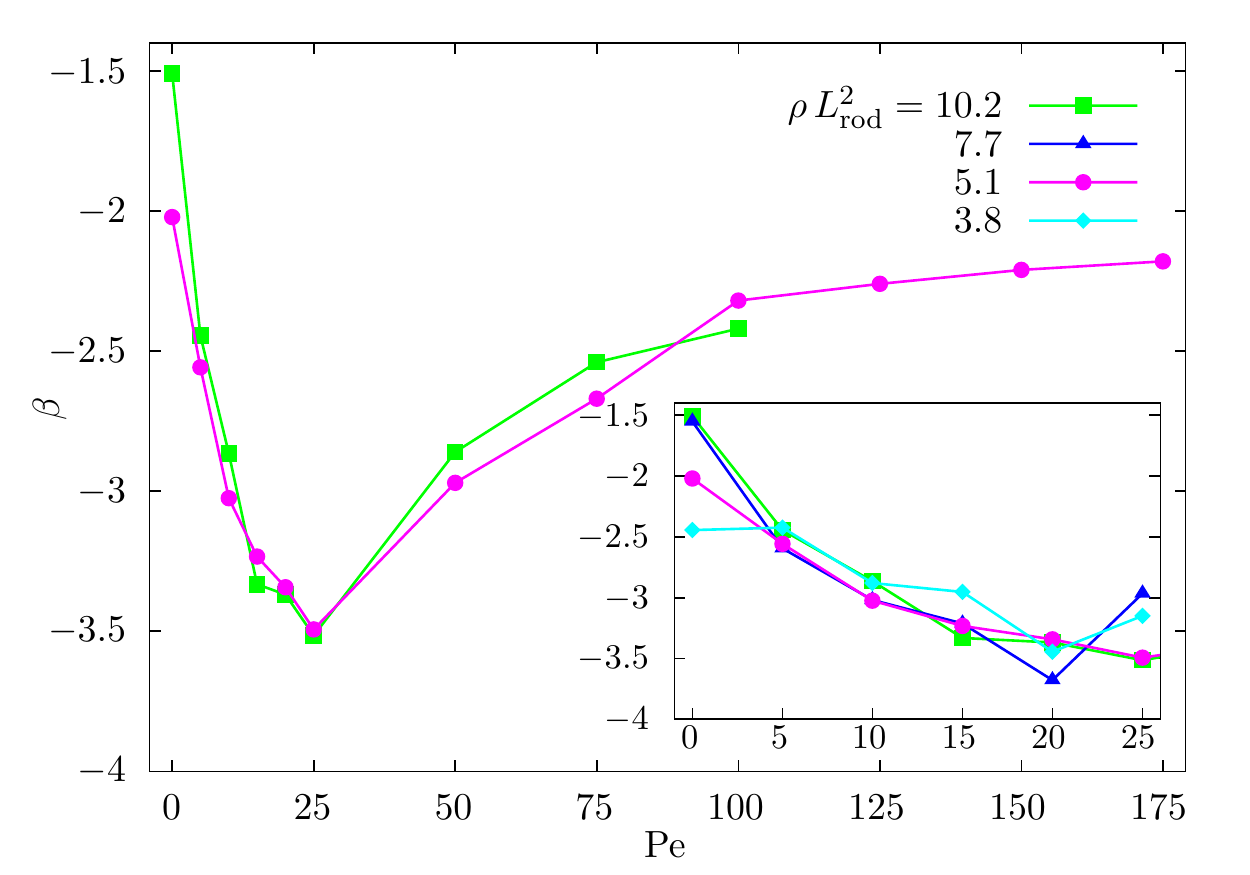}
\caption{(Color online) The exponent $\beta$ of the power law for cluster size distributions as a function of $\Pe$ for systems with $E=1.5\,\kBT$ and several average rod densities $\rho$. The exponents have very weak dependence on $\rho$. Inset: magnified view for $0<\Pe<25$.}
\label{fig:csdexponent}
\end{figure}
We define two rods to be in the same cluster if the nearest distance between
them is less than $2r_\mathrm{min}$ and the difference in their orientation
angles is less than $\pi/6$.
In Fig.~\ref{fig:sslj178csd}, sample cluster size distributions $\Pi(n)$ are presented. For small cluster size $n$, $\Pi(n)$ decreases with a power law, $\Pi(n)\propto n^{\beta}$ with $\beta<0$; for large $n$, $\Pi(n)$ decreases exponentially \cite{Yang2010,Huepe2008}.
For systems with giant clusters, such as the system with $\Pe=25$, there is a gap in the distribution because they consist of one giant cluster $(n>10\,000)$ and small clusters $(n<30)$ that mostly form near the boundary of the giant cluster. In such systems, the exponent $\beta$ is calculated only based on the distribution of small clusters.

\begin{figure}
\centering
  \includegraphics[trim= -0.0cm 0.0cm 0.0cm 0cm, clip=true, totalheight=0.71\columnwidth]{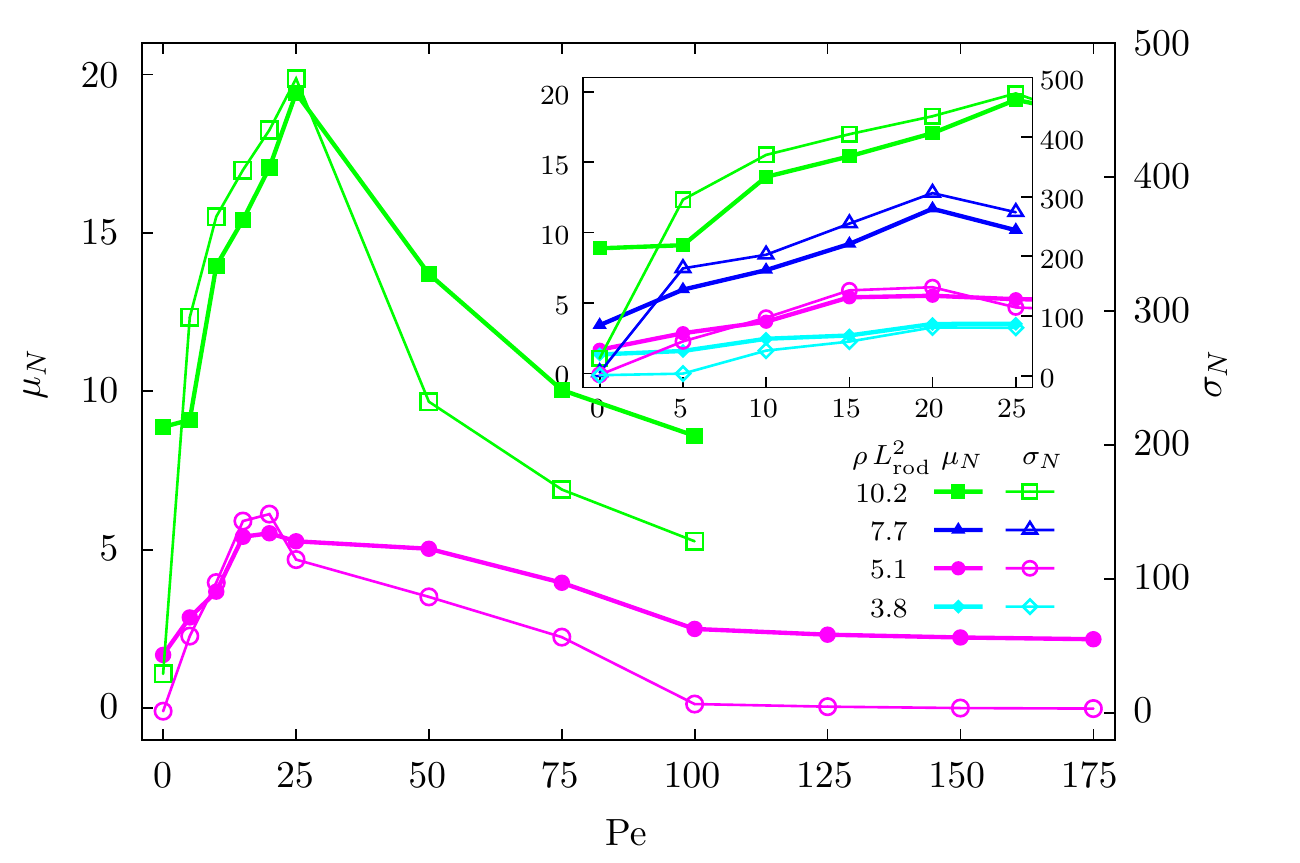}
\caption{(Color online) Cluster size average $\mu_N$ and spread $\sigma_N$ as function of $\Pe$ for several average system densities. The number of rods in systems with $\rho\,L_\rod^2=10.2$ is $13107$. Inset: magnified view for $0\le \Pe\le 25$. The cluster sizes have been averaged over $200$ frames in the last $40\,000$ time steps.}
\label{fig:avgclsize}
\end{figure}

The power-law exponent for the cluster size distribution first decreases with
increasing P\'eclet number, has a minimum for $\Pe\sim 25$, and then increases
for increasing the values of $\Pe$, see Fig.~\ref{fig:csdexponent}. We find
the exponent to be in the range $-1.5 \ge \beta \ge -3.5$, which agrees
with the range $-2 \ge \beta \ge -3.6$ found in Ref.~\cite{Yang2010} for
rods with different aspect ratio and a different interaction potential than
in our simulations. A recent experimental study found $\beta=-1.88\pm 0.07$
for clusters of \textit{M. xanthus} bacteria \cite{Peruani2012}.
As shown in Fig.~\ref{fig:avgclsize}, the average size of the clusters,
$\mu_N$, increases with increasing P\'eclet number for $\Pe\lesssim 25$ and
decreases if $\Pe$ is further increased. The spread of the cluster size,
$\sigma_N$, shows the same qualitative behavior but decays faster at
high $\Pe$ values, which shows that the system becomes more
homogeneous.

\subsection{Polar autocorrelation functions}
\label{sec:oacf}

\begin{figure}[!t]
\centering
  (a)\includegraphics[trim= 0.0cm 0.0cm 0.0cm 0cm, clip=true, totalheight=0.71\columnwidth]{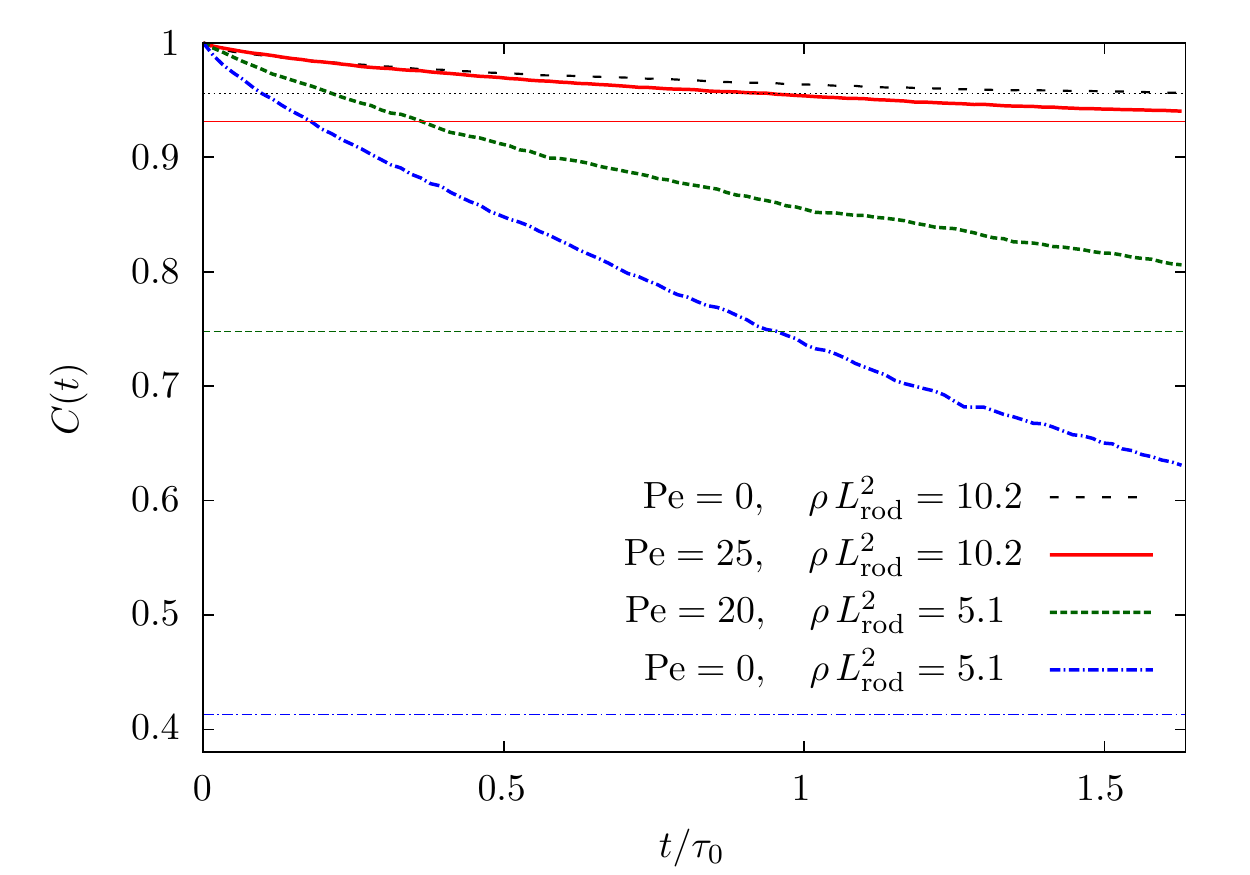} \\
  (b)\includegraphics[trim= 0.0cm 0.0cm 0.0cm 0cm, clip=true, totalheight=0.71\columnwidth]{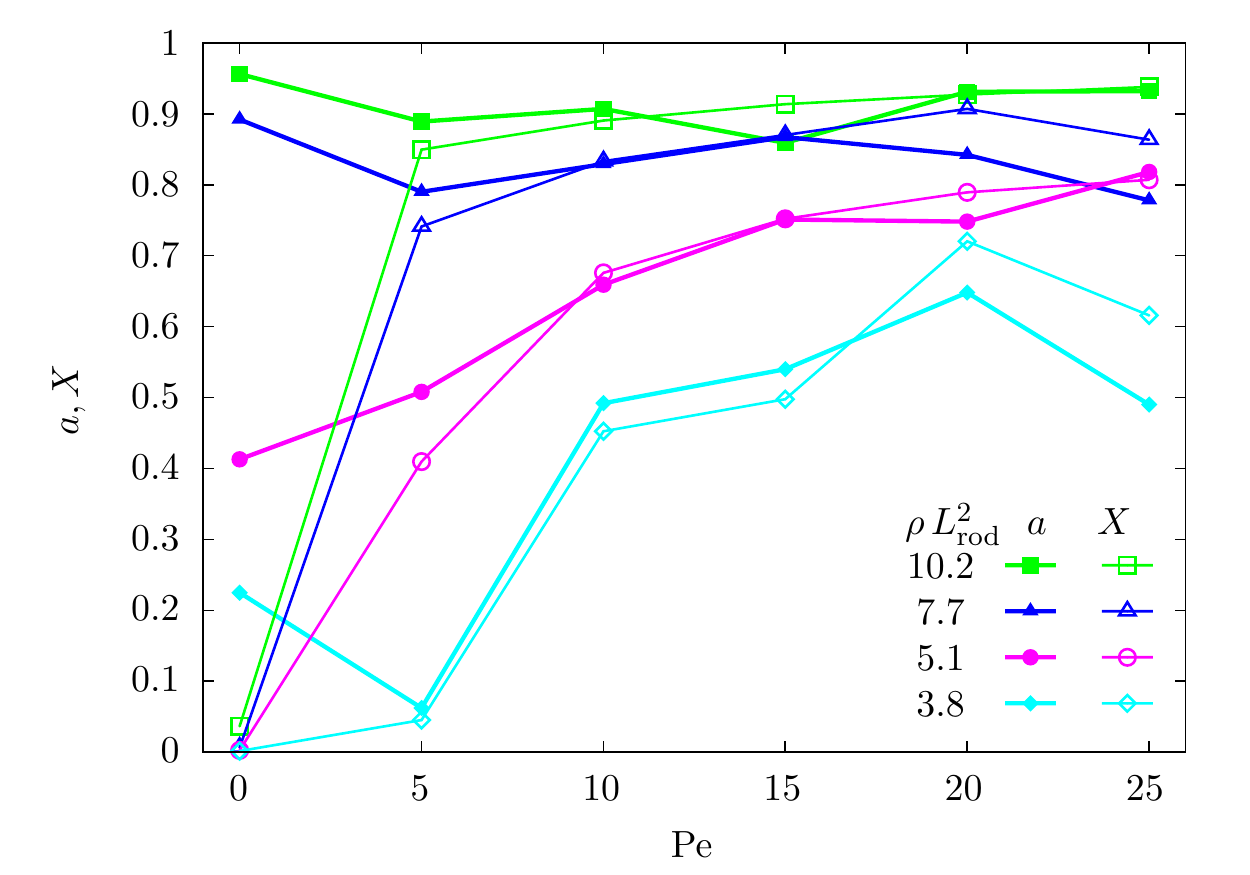} \\
  (c)\includegraphics[trim= 0.0cm 0.0cm 0.0cm 0cm, clip=true, totalheight=0.71\columnwidth]{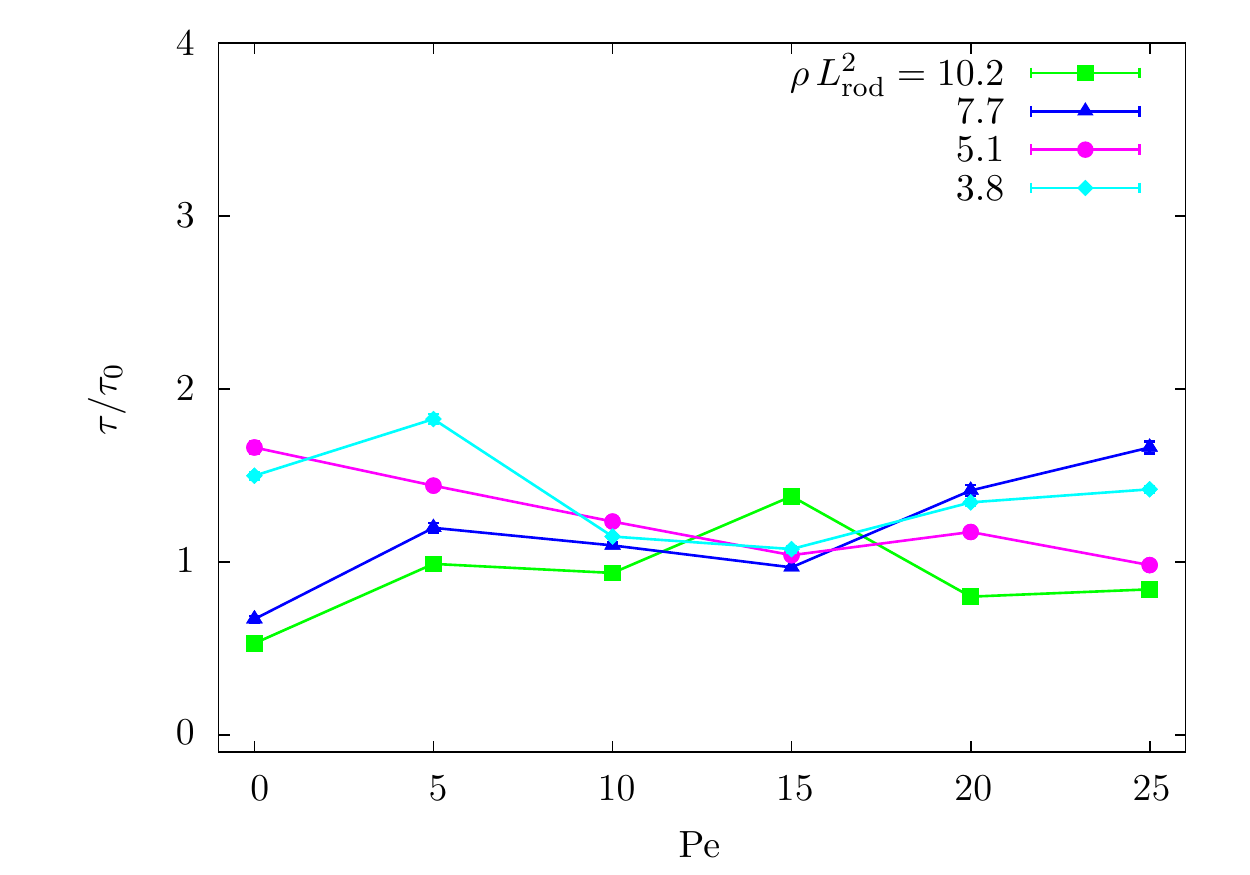}
\caption{(Color online) (a) Autocorrelation of rod orientation with lag time $t$ for the systems shown in Fig.~\ref{fig:snapshots}. Thick lines are simulation results; thin horizontal lines are autocorrelation base values, calculated by fitting the data with Eq.~(\ref{eq:oacf}). (b) Comparison of the autocorrelation base value $a$ and the fraction of rods in the largest cluster $X$. (c) Autocorrelation time $\tau$ of the rod orientation as function of $\Pe$ for several average rod densities. $\tau_0 = 1/D_r$ is the time unit; see Sec.~\ref{sec:model}.
The observables have been calculated based on $200$ frames in the last $40\,000$ time steps.
}\label{fig:OACFcombined}
\end{figure}

The clustering dynamics in the systems can be characterized by autocorrelation functions for the rod orientation
\begin{equation}
C(t)=\left< \mathbf{n}_i(t')\cdot\mathbf{n}_i(t'+t) \right>\,,
\end{equation}
for lag time $t$, where $\mathbf{n}_i(t')$ is the orientation vector of rod
$i$ at time $t'$, and the average is over all rods and over all times $t'$. 
Figure~\ref{fig:OACFcombined}(a) shows $C(t)$ for systems shown in 
Fig.~\ref{fig:snapshots}. The autocorrelation function $C(t)$
can be fit using a shifted exponential function
\begin{equation}
A(t)=(1-a)e^{-t/\tau}+a\,,
\label{eq:oacf}\end{equation}
where $\tau$ is the autocorrelation time and $a$ is an autocorrelation base value. 
A finite value of $a$ is the ratio of rods that do not lose their orientation for the time scale of the measurement.
Rods that are inside clusters are less likely to lose their orientation, which corresponds to a high value of $a$, while free rods in the gas change orientation more frequently because of rotational diffusion.

In Fig.~\ref{fig:OACFcombined}(b), we compare $a$ to the averaged fraction of
rods $X$ that are part of the largest cluster in the system for several
densities and P\'eclet numbers. In general, we find good agreement between
$a$ and $X$ \footnote{The values of $a$ and $X$ do not agree for $\Pe=0$, where rods are either freely moving
due to noise (isotropic regime) or stuck in small nonmotile clusters
(nematic regime). In the former case, there is hardly any orientation
preservation ($a\ll 1$) and there is no large cluster ($X\ll 1$). In the
latter case, rods hardly change their orientation ($a\sim 1$) but are
distributed over small clusters ($X\ll 1$).}.
The autocorrelation time $\tau$, shown in Fig.~\ref{fig:OACFcombined}(c),
does not change substantially for different values of $\Pe$ and $\rho$.
The correlation time $\tau$ obtained from the fit with Eq.~(\ref{eq:oacf})
is very similar to the autocorrelation time $\tau_0$ for a single rod,
which shows that the rotational diffusion is only weakly affected by
occasional collisions of the rods.
Therefore, the giant cluster moves persistently within simulation time.

\section{Summary and Conclusions}
\label{sec:summary}

We have studied collective behavior for self-propelled rigid rods in two
dimensions constructed by single beads that interact with a separation-shifted
Lennard Jones potential. The finite potential strength mimics the ability
of microswimmers close to a wall and of
filaments in motility assays to temporarily escape to the third dimension
and cross each other. 
For a high potential barrier, we recover the limit of impenetrable rods studied, for example, in Refs.~\cite{Yang2010,Wensink2012}.
For most simulations, we have used an interaction energy
$E=1.5\,\kBT$ that for complete overlap of two beads is of the order of the thermal energy; crossing of rods therefore occurs with a high probability; see Fig.~\ref{fig:CrossingRates}. 
However, the interaction energy is much larger than the bead-bead interaction energy if rods cross at a small angle or if a single rod approaches a cluster.
For our system with $n_{\rm b}=18$ beads per rod, the energy for complete
overlap is $n_b E = 27\,\kBT$ and thus the probability for such events is very low.

We have calculated a phase diagram for rod density and energy barrier to characterize the isotropic-nematic transition for passive rods.
The isotropic-nematic transition is shifted to higher densities for reduced overlap energy, because of the reduced rod-rod interaction. 
We find significant deviations from the transition density calculated for hard rods \cite{Onsager1944,Kayser1978} if the bead-bead interaction energy is below $2\,\kBT$.  
For $E=1.5\,\kBT$, the isotropic-nematic transition occurs for $\rho= 1.3\, \rho_c $, where $\rho_c$ is the transition density for hard rods \cite{Kayser1978}.
Our results using a modified Onsager theory show excellent agreement with our Monte Carlo simulations.

Using Brownian dynamics simulations, we have determined the crossing probability for two colliding rods as function of their relative angles for several values of penetrability coefficient and P\'eclet number. 
The crossing probability is highest for almost perpendicular collisions, which is qualitatively similar to the crossing probability measured in experiments with microtubules propelled on surfaces. 
In Ref.~\cite{Sumino2012}, the maximum crossing probability for two microtubules in a motility assay is $40 \, \%$ and corresponds to $Q=5$ and $\Pe=10$ in our simulations 
\footnote{The same crossing probability may be achieved by increasing $E$ and $\Pe$ at the same time.}.

Self-propelled rods align due to their soft repulsive interaction \cite{Baskaran2012}.
For high rod densities, we find a laning phase. 
For intermediate rod densities and P\'eclet numbers, we observe the formation of giant clusters that span the entire simulation box, which we denote as ``clustering window."
Clusters break if the propulsion force is strong enough to overcome the repulsive force due to rod-rod interaction.
We find a critical value $Q^*=28$ for cluster break-up.
We characterize our systems by cluster size distributions that can be fit by power-laws $\Pi(n)\propto n^\beta$ with $-1.5 \ge \beta \ge -3.5$, which is consistent with previous experimental and simulation results \cite{Yang2010,Peruani2012}.
By analyzing the autocorrelation function for rod orientation, we can separate the contributions from rods in a cluster from the contributions
from free rods. We find that the free rods show almost the same orientational correlation as single rods.

We can analytically estimate the density of free rods in systems with giant clusters, which we denote as ``gas density," $\rho_\gas = 192/(\pi^2 L_{\rm rod}^2 \Pe)$. 
The gas density is independent of the average rod density in the system, which is analogous to the molecule density in the gas phase for liquid-gas coexistence that does not depend on the volume of the liquid phase but only on temperature. 
Using $E_b = \kBT \ln [192 r_{\rm min}/(\pi^2 L_{\rm rod} \Pe)]$, we calculate effective binding energies for rods in the cluster. 
For aspect ratio $18$ used in our simulations and $\Pe \approx 25$, we find effective binding energies of about $0.1\,\kBT$, which is comparable to binding energies for the gas-liquid critical point for colloidal systems \cite{Vliegenthart2000}.

Phase separation into high-density and low-density regions is an intrinsic property of self-propelled particle systems and has also been observed for nonaligning spherical particles \cite{Tailleur0805,fily1206,redner1301,Wysocki1308}. 
As for the rods, the gas density of the spheres is inversely proportional to the propulsion velocity \cite{redner1301}.
However, the nature of cluster formation is different in the two models: While we observe motile clusters as a result of particle alignment, systems with nonaligning spheres exhibit jammed \textit{non}motile clusters as a result of steric trapping.
Moreover, the internal structure of clusters is nematic in our model, contrary to the isotropic structure for nonaligning spheres.
Of course, also laning phases are only possible for anisotropic particles.

We have introduced and characterized a model of self-propelled rods that
interact with a physical interaction that allows for crossing events. The
model can now be used to interpret experiments for almost two-dimensional
systems with good computational efficiency and allows predictions beyond those based on models using point
particles with phenomenological alignment rules.

\acknowledgments
We thank Adam Wysocki and Jens Elgeti for stimulating discussions. M.A. and K.M. acknowledge support by the International Helmholtz Research School of Biophysics and Soft Matter (IHRS BioSoft). CPU time allowance from the J\"ulich Supercomputing Centre (JSC) is gratefully acknowledged.

\appendix
\section{Phase transition of passive rods}
\label{app:passivetransitions}

\begin{figure}[!ht]
\centering
  (a)\includegraphics[trim= 0.0cm 0.0cm 0.0cm 0cm, clip=true, totalheight=0.71\columnwidth]{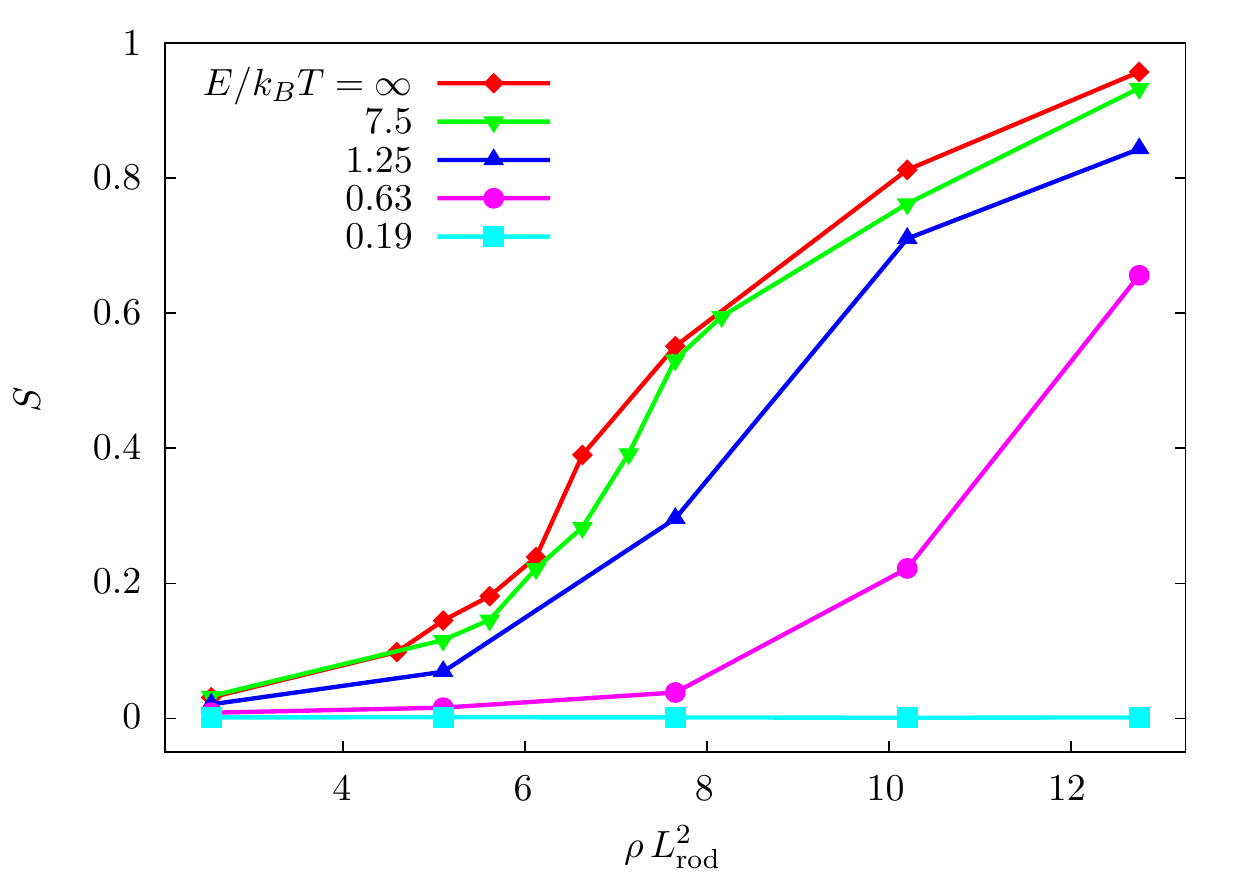} \\
  (b)\includegraphics[trim= 0.0cm 0.0cm 0.0cm 0cm, clip=true, totalheight=0.71\columnwidth]{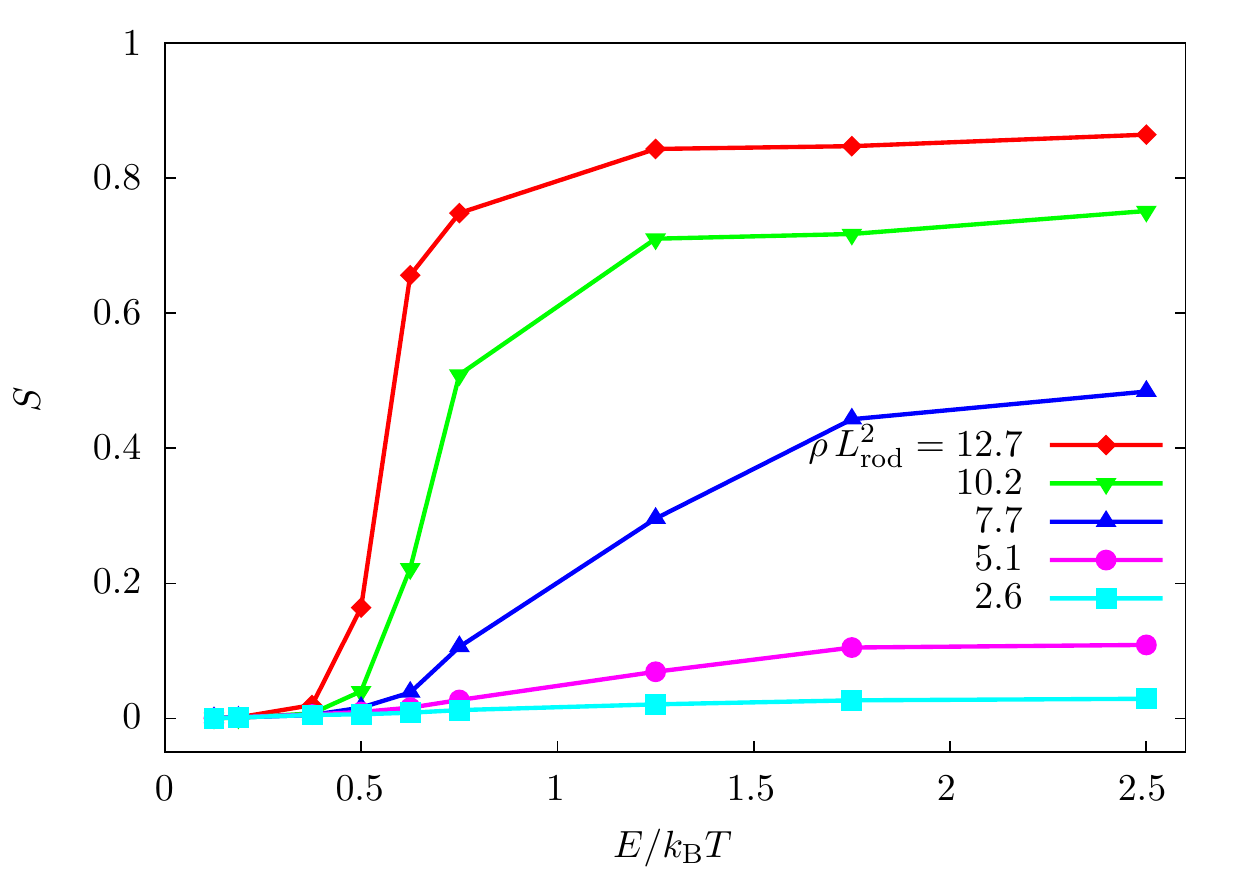}
\caption{(Color online) Nematic order parameter $S$, used to determine the phase transition for passive rods, as a function of (a) the rod density $\rho$ for several energy barriers $E$ and (b) the energy barrier $E$ for several rod densities $\rho$.}\label{fig:PhaseTransitionCombined}
\end{figure}

The nematic order parameter is
plotted in Fig.~\ref{fig:PhaseTransitionCombined} for various cuts through the phase diagram in Fig.~\ref{fig:IsoNemaPhaseDiag}.
It has been suggested that the isotropic-nematic transition of rods is continuous in two dimensions \cite{Kayser1978}.
We have chosen a threshold value for the isotropic-nematic phase transition,
$S_t=0.11$, such that for an infinite interaction energy the value predicted by the
Onsager theory is recovered. Our threshold value is similar to the threshold
value $S=0.2$ that has been chosen in Ref.~\cite{Kraikivski2006}.

For finite energy barrier $E$, we generalize the approach presented in
Ref.~\cite{Kayser1978} based on bifurcation theory to obtain the critical
density for the isotropic-nematic transition.
The distribution function for the rod orientation is given by $f(\theta)$,
which satisfies 
\begin{equation}
\label{eq:lnf}
\ln[2\pi f(\theta)] = C + \lambda\int_{\theta=0}^{2\pi} F(\theta,\theta')f(\theta')\ud\theta'\,,
\end{equation}
where the constant $C$ is determined by the normalization of $f(\theta)$,
\begin{equation}
\label{eq:fnorm}
\int_{\theta=0}^{2\pi}f(\theta)\ud\theta=1\,.
\end{equation}
We define $h(\theta)$ as
\begin{equation}
\label{eq:hdef}
h(\theta)=2\pi f(\theta)-1\,,
\end{equation}
such that $h(\theta)=0$ corresponds to an isotropic distribution. 
Using Eqs.~(\ref{eq:lnf})--(\ref{eq:hdef}), we can write
\begin{equation}
h(\theta)=-1+\frac{\exp(\lambda\int F(\theta,\theta')h(\theta')\ud\theta'/2\pi)}{(1/2\pi) \int\exp(\lambda\int F(\theta,\theta')h(\theta')\ud\theta'/2\pi)\ud\theta}\,.
\label{eq:Htheta1}
\end{equation}

We assume that the interaction energy of two rods is either $0$ or $E$, depending on whether they cross each other. This approximation is justified if the rods are very thin and the complete overlap of two rods---which is energetically very unfavorable---is excluded.
In the regime where $L_\rod \gg r_{\rm min}$, the parameter $\lambda$ and the kernel $F$ are given by 
\begin{eqnarray}
\lambda &=& \frac{1}{4}\rho\pi L_\rod^2\label{eq:lambda}\,,\\
(1/2\pi)F(\theta,\theta') &=& (-2/\pi^2)\sin(\theta) \label{eq:F} \\
& &\times [1-\exp(-E/\kBT)] \, . \nonumber
\end{eqnarray}
Substituting Eqs.~(\ref{eq:lambda}) and (\ref{eq:F}) in Eq.~(\ref{eq:Htheta1}) gives
\begin{equation}
h(\theta)= -1 + \frac{\exp[-\rho K h(\theta)]}{(1/\pi)\int_0^{\pi}\exp[-\rho K h(\theta)]\ud\theta}\,,
\end{equation}
where the operator $K$ is defined as 
\begin{eqnarray}
K h(\theta) &=& (L_\rod^2/\pi)[1-\exp(-E/\kBT)] \\
& &\times\int_{0}^{\pi} \sin(\theta')h(\theta-\theta')\ud \theta'\,.\nonumber
\end{eqnarray}
For $h(\theta)$ to have bifurcation point in $\rho$, 
\begin{equation}
w(\theta) = -\rho K w(\theta)\label{eq:linear}\,
\end{equation}
has to have an eigenfunction with two maxima at $w(0)=w(\pi)$ and no further
maxima. The corresponding eigenvalue determines the density at which
bifurcation occurs.
The desired eigenfunction is $\cos 2\theta$ with the eigenvalue $-3/2$;
thus the bifurcation density that corresponds to the isotropic-nematic
transition is 
\begin{equation}
\rho_c=\frac{3\pi}{2L_\rod^2} \frac{1}{[1-\exp(-E/\kBT)]}\,.
\label{eq:IsoNem2}
\end{equation}

\section{Effective binding energy for rod adsorption to the cluster}
\label{app:bindingenergy}
The independence of gas density $\rho_\gas$ from the average density of the system $\rho$ is analogous to a vapor density for rods. 
Here we follow this analogy to obtain an effective binding energy gain $E_b$ for rods that are part of the cluster.

We use an ideal-gas model in two dimensions to represent the rods in the gas phase. The activity and the anisotropy of the rods are intentionally not taken into account explicitly and enter via the effective binding energy.
The free energy for the rods in the gas is thus
\begin{eqnarray}
F_\gas &=& N\kBT \ln\left(\frac{N \Delta}{A}\right) \\
 &=& N\kBT \ln\left(\rho_\gas L_\rod r_{\rm min}\right) \nonumber\,,
\end{eqnarray}
where $N$ is the number of rods in the gas, $\Delta$ is the area of each rod, and $A$ is the area accessible for the rods in the gas.

In the cluster, each rod gains a binding energy $E_b$,
\begin{equation}
F_{\rm cluster}=N E_b \, ,
\end{equation}
where $N$ here is the number of rods in the cluster. In equilibrium, the chemical potential $\mu = \partial F/\partial N $ in the gas and in the cluster should be equal. This gives Eq.~(\ref{eq:bindingenergy}).

\bibliographystyle{apsrev4-1}
\bibliography{references}

\end{document}